\documentclass{aa}  

\usepackage{graphicx}
\usepackage{placeins}
\usepackage{xcolor}
\usepackage{mathtools}
\usepackage{bbm}
\usepackage{placeins}
\usepackage{url}
\usepackage{amsmath}
\usepackage{multirow}
\usepackage{txfonts}

\newcommand*{\defeq}{\mathrel{\vcenter{\baselineskip0.5ex \lineskiplimit0pt
                     \hbox{\scriptsize.}\hbox{\scriptsize.}}}
                     =}

\begin{document}

%   \title{Computing magnitudes from low signal-to-noise observations}
   \title{Computing magnitudes, colours, distances, and absolute magnitudes at any signal-to-noise level}

   \author{M. Weiler\inst{1,2,3}
             }

\authorrunning{M. Weiler}

   \institute{Departament de F{\'i}sica Qu{\`a}ntica i Astrof{\'i}sica (FQA), Universitat de Barcelona (UB), c. Mart{\'i} i Franqu{\`e}s, 1, 08028 Barcelona, Spain
   \and
   Institut de Ci{\`e}ncies del Cosmos (ICCUB), Universitat de Barcelona (UB), c. Mart{\'i} i Franqu{\`e}s, 1, 08028 Barcelona, Spain
   \and
   Institut d'Estudis Espacials de Catalunya (IEEC), Esteve Terradas 1, Edificio RDIT, Ofic. 212, 08860 Castelldefels, Spain\\
              \email{mweiler@fqa.ub.edu}
             }

   \date{Received 27 March 2025; accepted ??}

  \abstract
    % context heading (optional)
   {The computation of both, magnitudes and distances from low signal-to-noise observations is known to be problematic, in the sense that the magnitudes and distances tend to assume extreme values, or are even undefined in the case of negative observed fluxes, or unphysical in the case of negative observed parallaxes.} 
  % aims heading (mandatory)
   {In this work we show that magnitudes can be computed consistently at all signal-to-noise levels, and even for negative fluxes, if the prior information that the true flux or distance is non-negative is properly included. Furthermore, we derive an ``all-purpose'' estimator for distances from a prior implementing only the non-negativity of the true parallax. We apply our results to the case of combining magnitudes to colours, and magnitudes and distances to obtain absolute magnitudes.}
  % methods heading (mandatory)
   {We introduce the priors in the computation of magnitudes and distances, and compute the resulting distribution functions for the true magnitudes and true distances. From these distribution functions we derive simple expressions for the estimators of the magnitude and distances, as well as their uncertainties, which can be consistently applied to observations at all signal-to-noise levels, and for all observed fluxes and parallaxes.}
  % results heading (mandatory)
   {The resulting expressions are easy to compute and we show that the resulting distribution functions for magnitudes, colours, distances, and absolute magnitudes are not only consistent for all signal-to-noise levels and applicable to both, positive and negative observed fluxes and parallaxes, but also show no strong tails. While biases at very low signal-to-noise levels are unavoidable, the estimator for distances derived in this work is less biased than previously used estimators. We find that the magnitude, colour, distance, and absolute magnitude distributions for vanishing signals converge to limiting distributions, whose median values are important for assessing biases when working with data at low signal-to-noise levels.}
{}
 

   \maketitle
%
%-------------------------------------------------------------------

%===================================
\section{Introduction \label{sec:introduction}}
%===================================

The brightness of an astronomical object is frequently expressed in terms of its magnitude, $m$, which is defined as
\begin{equation}
m = -2.5 \cdot {\rm log_{10}}(\phi) + zp \, . \label{eq:def}
\end{equation}
In this equation, $zp$ is a zero point, and $\phi$ is ``the quantity of light'' recorded within a passband filter. This definition, though looking simple, comes with a number of issues, which already begin with what the ``quantity of light'' is. If one uses a photon-counting instrument (such as a CCD detector or any other kind of detector making use of the photoelectric effect), then the amount of light is a number of photons recorded per unit of time and detector area. If one uses an energy-integrating detector (such as a bolometer), then the quantity of light is an amount of energy recorded per unit of time and detector area. If the passband filter in which the photometric observations are made has a non-negligible width, then both quantities are different and cannot be converted into each other. The reason is that it is unknown if a recorded number of photons originated more on the short-wavelength part of the passband, and thus carried more energy, or from the long-wavelength part of the passband, and therefore carried less energy. And vice versa, having recorded a certain amount of energy, it is unknown if this energy was deposed in the detector a smaller number of higher energy photons, or a larger number of low-energy photons.\par
The choice of the zero point is not unique, and the most used choices tie the zero point to the star Vega ($\alpha$ Lyr), or, in the AB photometric system, to a spectrum with constant flux in frequency \citep{Oke1983}. The factor $-2.5$ in Eq.~(\ref{eq:def}) goes back to attaching the brightness system to a system of six classes of brightnesses for stars used by Ptolemy in the \textit{Almagest} \citep{Hearnshaw1996}, and the logarithmic dependency of the magnitudes reflect the logarithmic sensitivity of the human eye. The use of the decadic instead of the natural logarithm probably goes back to the fact that humans have ten fingers available for counting.\par
A further issue with this definition occurs when applying it to observed fluxes $\phi$ which are significantly affected by errors, i.e. to observations with low signal-to-noise ratios. If the observed flux becomes negative, which may happen at low flux levels after subtraction of noisy background components, the magnitude is undefined. And the magnitude values can take extreme values if the observed flux is positive, but strongly affected by noise. It is therefore in general advisable to work with fluxes instead of magnitudes. The use of magnitudes is however widely common, and to improve the use of magnitudes at low signal-to-noise levels, \cite{Lupton1999} have proposed a compromise. They suggested to replace the logarithmic dependency in the magnitude definition by a dependency on the inverse hyperbolic sine function, such that for high signal-to-noise values the dependency approximates the usual logarithmic behaviour, while at low absolute values of the measured signal-to-noise ratios the dependency becomes approximately linear. While this approach is viable, it is certainly an ad-hoc solution and adds another layer of arbitrariness to the definition of magnitudes. In this work we show that no ad-hoc modifications are required, and that magnitudes can be computed consistently at all signal-to-noise levels, including negative observed fluxes, if the prior information that the true flux of an astronomical object is non-negative is included into the computation of the magnitude of the object.\par
Also the computation of distances from parallaxes is complicated by a number of issues, and \cite{Luri2018} provide an overview of the complications in converting observed parallaxes into distances. One of these complications is that observational parallaxes might become negative at small signal to noise ratios, which represents an unphysical result. The true parallax, not affected by noise, has to be non-negative. The situation is thus related to the problem of converting fluxes into magnitudes, both on a practical and a theoretical level. On a practical level, one important application of distances is the combination with a magnitude to obtain the absolute magnitude of an object. On a theoretical level, both problems consist of applying a continuous, strictly monotonic non-linear transformation to a noisy quantity which a-priori is non-negative. In the case of magnitudes, this transformation is the logarithmic transformation of Eq.~(\ref{eq:def}). In case of distances the transformation is the inversion of the parallax, $\varpi$, measured in arcseconds, into a distance $D$, measured in parsec,
\begin{equation}
D = \frac{1}{\varpi} \; . \label{eq:defDist}
\end{equation}
In this work we show that both problems, the computation of magnitudes from noisy fluxes and the computation of distances from noisy parallaxes, can be solved in the same basic way, as long as one is interested in a universally applicable estimator that implements the non-negativity of the true physical quantities.

%===================================
\section{Measurements and probability \label{sec:measurements}}
%===================================

In order to include the prior information that the true value is non-negative into the computation of magnitudes or distances, one has to think of a measurement in terms of a probability density function (pdf). One usually thinks of the result of a measurement as a value and its error. This view however is simplified, as the actual result of a measurement is a pdf for the values measured by the instrument. Often this pdf is assumed to be the one of a normal distribution, which then is characterised by its mean, the ``value'', and its standard deviation, the ``error''. If another quantity is computed from the measured one, and the transformation preserves at least approximately the normal distribution, then the pdf of the resulting quantity is again characterised by the mean and the standard deviation, and all computations can be done with these two quantities alone. This reinforces the notion of working with values and their errors. If however strongly non-normal pdfs are involved, or the transformation of the measured quantity does not preserve the normal distribution of the original measurement, then the mean and the standard deviation are no longer good measures to describe the resulting pdfs. In such cases the full pdf, as the actual outcome of a measurement, has to be transformed into the pdf of the derived quantity. Both aspects, non-linear transformations that do not preserve the normality of the pdf, and intrinsically non-normal pdfs, play a role in the problem of computing magnitudes from fluxes, and distances from parallaxes, at low signal-to-noise values.\par
Another important aspect of the computation of magnitudes and distances at low signal-to-noise values is a subtle but critical change in interpretation, namely the step from ``the measured flux or parallax is this value with this error'' to ``the flux or parallax of the observed object is this value with this error''. This transfer from the probability of the observed value to the probability of the true value assumes that the probability of the measurement, given the true value for the object, is the same as the probability of the true value for the object, given the measurement. This equality is correct according to Bayes' theorem only in case of a non-informative prior, i.e. in cases when all outcomes are equally probable a-priori. This however is not the case in photometric measurements or distance determinations, as there is the prior information that the true flux or distance must be non-negative. Using this prior is not necessary at large signal-to-noise ratios, as the normal distribution falls off so quickly that the probability of negative values is already virtually zero, and it is therefore not required to use the prior to set it explicitly to zero. At low signal-to-noise ratios, the prior however needs to be explicitly taken into account when deriving the probability of the true value from the probability of the measured value. The core problem with applying Eq.~(\ref{eq:def}) or (\ref{eq:defDist}) to low signal-to-noise regimes is thus that the probability of the true value, given the measured value, is no longer equal to the probability of measured value, given the true value: at low signal-to-noise ratios, the probability for measuring a negative flux or parallax of some object, given its (non-negative) true flux or parallax, becomes significantly larger than zero, while the probability of the true, negative flux or parallax, given a negative measured flux or parallax, remains exactly zero. When taking this effect into account, and by transforming the resulting pdfs from the flux to the magnitude domain, a consistent computation of a magnitude or distance for any measured flux or parallax and the intervals of confidence becomes easily possible.

%===================================
\section{The magnitude distribution \label{sec:magnitudes}}
%===================================

We assume here that the pdf of the observed flux is the one of a normal distribution, with mean $\mu_\phi$ and standard deviation $\sigma_\phi$. This assumption is an approximation, as the photon noise of the background and the flux follow a Poisson distribution. However, summing the independent random noise components, such as the Poisson-distributed contribution from the background and the observed object and the noise contribution from the detector, might result in a distribution that is well approximated by a normal distribution even for low flux levels. The pdf of the true flux is the normalised product of the pdf of the observed flux and the prior. We set the prior probability to zero for negative fluxes, and assume it to be constant for non-negative fluxes. We may further assume a priori that all measured fluxes are finite. In this case, the prior can be normalised. The upper bound of the prior is not of relevance as long as it is large compared to the flux level on which the pdf of the flux is significantly different from zero. In this case, the resulting pdf for the true flux is a truncated normal distribution.\par
If $\Phi(x)$ denotes the cumulative distribution function of the standard normal distribution,
\begin{equation}
\Phi(x) = \frac{1}{2} \left[ 1 + {\rm erf}\left( \frac{x}{\sqrt{2}}\right) \right] \, ,
\end{equation}
then the pdf of a truncated normal distribution, truncated to an interval $[A, \infty]$, is
\begin{equation}
f_{\boldsymbol{T}}(x) = \frac{1}{\sqrt{2\pi}\, \sigma} \, \frac{1}{1 - \Phi\left( \frac{A-\mu}{\sigma} \right)} \, {\rm e}^{-\frac{(x-\mu)^2}{2\, \sigma^2}} \, \text{for }x \ge A, \; f_{\boldsymbol{T}}(x)=0 \, \text{else.}
\end{equation}
And while the pdf of the \textit{observed} flux has support for negative values and thus cannot be transformed logarithmically, the pdf for the \textit{true} flux, with $A \ge 0$, can be transformed logarithmically. The truncated normal distribution however becomes very much non-normal if the truncation becomes significant, so it is necessary to explicitly transform it in order to find the pdf for the true magnitude.\par
In this work, we write the pdf of a random variable as $f_{\boldsymbol{x}}(x)$, and the corresponding cumulative distribution as $F_{\boldsymbol{x}}(x)$. The bold face is used to distinguish between the name of the random variable and its value. If a strictly monotonic function $g(x)$ is applied to a random variable which is distributed according the $f_{\boldsymbol{x}}(x)$, then the pdf of the transformed variable $y=g(x)$ is given by \citep{Borovkov2013}
\begin{equation}
f_{\boldsymbol{y}}(y) = f_{\boldsymbol{x}}\left( g^{-1}(y)\right) \cdot \left| \frac{{\rm d}}{{\rm d}y}g^{-1}(y) \, \right| \, . \label{eq:generalTransform}
\end{equation}
With $g(x)$ the magnitude transformation as defined in Eq. (\ref{eq:def}), the pdf of the general magnitude distribution becomes thus
\begin{equation}
f_{\boldsymbol{m}}(m) = \frac{{\rm ln}(10)}{2.5} \, 10^{0.4\, (zp - m)} \, f_{\boldsymbol{\phi}}\left( 10^{0.4\,(zp-m)}\right) \, .
\end{equation}
Taking the pdf for the true flux to be the one of a truncated normal distribution, with $A=0$ \footnote{The choice of $A=0$ implies that an object is observed independently of its flux level. If a detection threshold applies, then $A$ might be larger than $0$. This avoids very low signal-to-noise levels but introduces similar effects, that are not within the scope of this work.}, one obtains the pdf of the distribution of the true magnitude,
\begin{equation}
f_{\boldsymbol{m}}(m) =  \sqrt{\frac{2}{\pi}} \, \frac{{\rm ln}(10)}{2.5} \frac{1}{\sigma_\phi} \, \frac{1}{1+ {\rm erf}\left( \frac{\mu_\phi}{\sqrt{2}\sigma_\phi} \right) } \, x \, {\rm e}^{-\frac{(x-\mu_\phi)^2}{2\sigma_\phi^2}}, \label{eq:magDistFunc}
\end{equation}
where $x = 10^{0.4\, (zp - m)}$.\par
The pdf of this distribution is strongly asymmetric for small signal-to-noise values, with a strong tail towards very large magnitudes, which is responsible for the frequent observation of extreme magnitudes at low signal-to-noise levels.\par
It is impractical to specify the full pdf of the distribution for the magnitude of a particular measurement. Just as normally distributed measurements are specified by the mean and the standard deviation of their pdf, one might wish to characterise a particular magnitude distribution by simple parameters that characterise its position and extend. The mean and the standard deviation however might not be suitable to characterise an asymmetric and strongly tailed distribution. Quantiles might be a better choice for this purpose, as they are both, easy to compute, and sensitive to the asymmetry of the function. For deriving the quantile function, we first compute the cumulative probability function of the magnitude distribution,
\begin{equation}
F_{\boldsymbol{m}}(m) = \int\limits_{-\infty}^m \, f_{\boldsymbol{m}}(m^\prime) \, {\rm d}m^\prime \, ,
\end{equation}
The integrals solved in this and the following sections of this work can with suitable substitutions essentially be reduced to forms, which are \citep{Gradshteyn2007}
\begin{eqnarray}
\int\, {\rm e}^{-a\, x^2} \, {\rm d}x & = & \sqrt{\frac{\pi}{4\, a}}\, {\rm erf}\left( \sqrt{a}\, x  \right) \, , \;  a > 0\\
\int\, x \, {\rm e}^{a\, x^2} \, {\rm d}x & = & \frac{{\rm e}^{a\, x^2}}{2\, a} \, \\
\int\, {\rm e}^{-\left(a\, x^2 + 2\,b\, x + c\right)}\, {\rm d}x & = & \frac{1}{2} \, \sqrt{\frac{\pi}{a}} \, {\rm e}^{\frac{b^2}{a} - c} \, {\rm erf}\left( \sqrt{a}\, x + \frac{b}{\sqrt{a}} \right) \, .
\end{eqnarray}
For the cumulative probability function we obtain\\
\begin{equation}
F_{\boldsymbol{m}}(m) = \frac{1}{1+ {\rm erf}\left( \frac{\mu_\phi}{\sqrt{2}\, \sigma_\phi}\right)} \, \left[  1 - {\rm erf}\left(  \frac{1}{\sqrt{2}\,\sigma_\phi} \, \left( 10^{0.4\, (zp - m)} - \mu_\phi \right)   \right) \, \right] \, .
\end{equation}
By inverting the cumulative probability function, we obtain the corresponding quantile function as
\begin{equation}
m(F_{\boldsymbol{m}}) = -2.5 \cdot {\rm log_{10}}\left( \mu_\phi + \sigma_\phi \cdot C\left( \frac{\mu_\phi}{\sigma_\phi},F_{\bf m} \right)  \right) + zp \label{eq:quantiles}
\end{equation}
where
\begin{equation}
 C(x,p) = \sqrt{2} \cdot {\rm erf}^{-1}\left( 1 - p \cdot \left[ 1 + {\rm erf}\left( \frac{x}{\sqrt{2}}\right) \right] \, \right) \label{eq:correctionFunction}
 \end{equation}
with $p \in (0,1]$, and ${\rm erf^{-1}}(x)$ denoting the inverse error function.\par
\begin{figure}
   %\centering
   \includegraphics[width=0.47\textwidth]{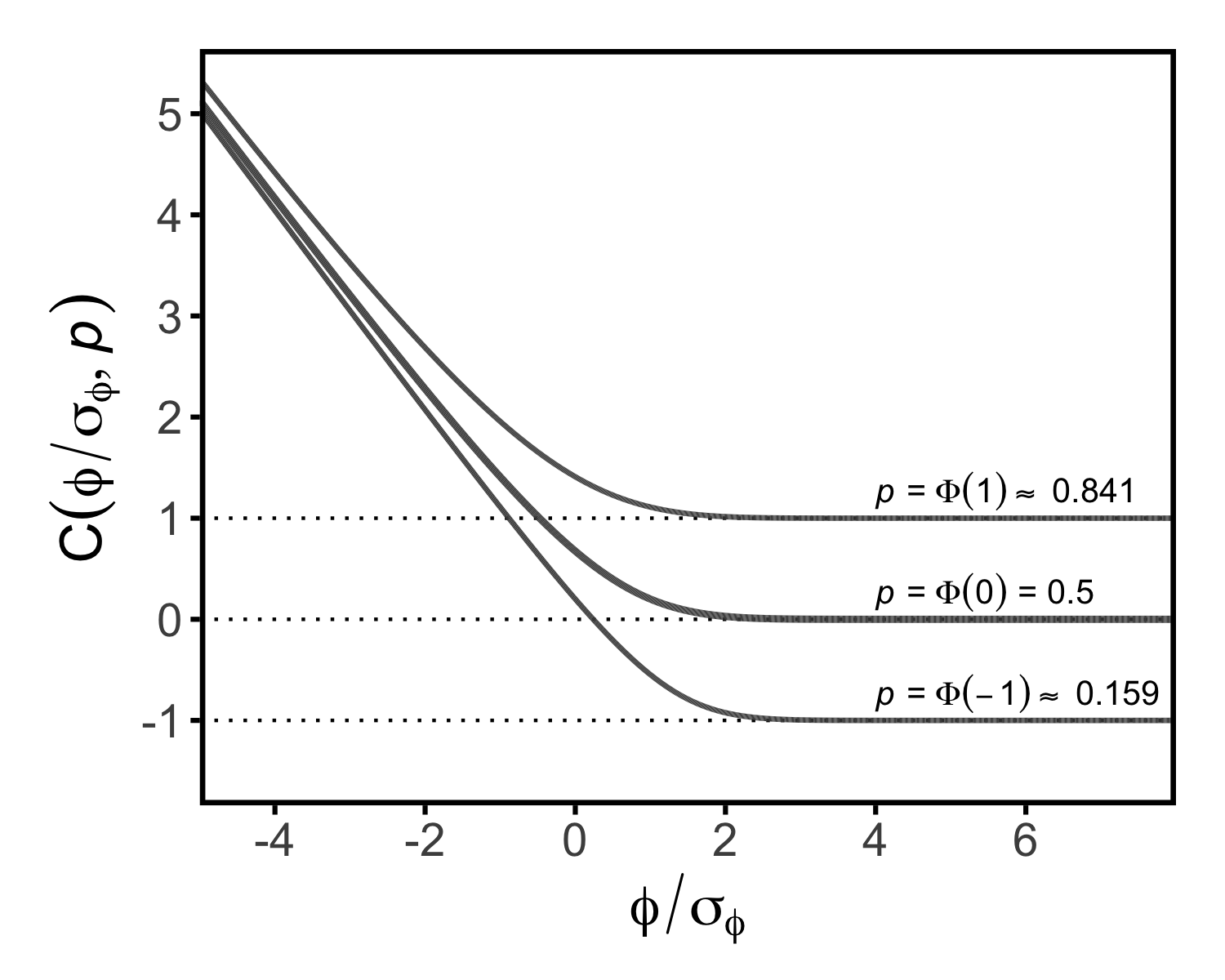}
   \caption{The correction function $C(x,p)$, with $x = \phi/\sigma_\phi$ the ratio of the observed flux and the flux error, for three different values of $p$.}
              \label{fig:correctionFunction}
    \end{figure}

This quantile function has essentially the form of the definition of magnitude as in Eq. (\ref{eq:def}), only that the standard deviation of the flux is added to the flux, weighted with the correction function $C(\mu_\phi/\sigma_\phi,F_{\boldsymbol{m}})$. Figure~\ref{fig:correctionFunction} shows the values of the correction function around small values of the signal-to-noise ratio, for three different values of $F_{\boldsymbol{m}}$, namely $\Phi(1)$, $\Phi(0)$, and $\Phi(-1)$. $\Phi(0) = 0.5$ corresponds to the median of the magnitude distribution function, and for increasing signal-to-noise ratios, the correction function for the median quickly converges to zero, thus turning Eq. (\ref{eq:quantiles}) into Eq.~(\ref{eq:def}). The quantiles for $\Phi(1) \approx 0.841$ and $\Phi(-1) \approx 0.159$ correspond to plus and minus one standard deviation in a normal distribution. The correction function for these two quantiles quickly converge to 1 and $-1$, respectively, for increasing signal-to-noise ratios, which corresponds to the magnitude for the mean flux plus and minus one standard deviation. These three quantiles might thus be a suitable choice for characterising the magnitude distribution function, as for high signal-to-noise ratios they quickly converge to the magnitude and its uncertainty computed in a naive way from Eq.~(\ref{eq:def}), while at low signal-to-noise ratios they reflect the asymmetry of the distribution function.\par
 We have now an easy to use estimator for the true magnitude and an its confidence intervals, $m^{+ \sigma^+}_{-\sigma^-}$, consistently from any observed flux $\phi \in \mathbb R$, as
\begin{equation}
m = -2.5 \cdot {\rm log_{10}}\left(\phi  + \sigma \cdot C\left(\frac{\phi}{\sigma_\phi},\frac{1}{2}\right)\right) + zp \, , \label{eq:newMag}
\end{equation}
and the lower limit as
\begin{equation}
\sigma^- = m + 2.5 \cdot {\rm log_{10}}\left(\phi  + \sigma_\phi \cdot C\left(\frac{\phi}{\sigma_\phi},\Phi(-1)\right)\right) + zp \, , \label{eq:newLow}
\end{equation}
and the upper limit
\begin{equation}
\sigma^+ = -2.5 \cdot {\rm log_{10}}\left(\phi  + \sigma_\phi \cdot C\left(\frac{\phi}{\sigma_\phi},\Phi(1)\right)\right) + zp - m\, . \label{eq:newHigh}
\end{equation}
If $\phi$ is the best guess for the mean of the observed flux, then $m$ according to Eq.~(\ref{eq:newMag}) is the best guess for the median magnitude, assuming a non-negative true flux.\par
One important property of the magnitude distribution function is the fact that it converges towards a limit distribution when the true flux of the source goes towards zero. This limit distribution solely depends on the noise, and it follows from setting $\mu = 0$ in Eq.~(\ref{eq:magDistFunc}). The median of the magnitude distribution function converges towards a limiting magnitude
\begin{equation}
m_{\rm limit} = -2.5 \cdot {\rm log_{10}}\left( \sigma \cdot C\left( 0,\frac{1}{2}\right) \right) + zp \, \label{eq:limMag}
\end{equation}
with $C\left( 0,\frac{1}{2}\right) = \sqrt{2} \, {\rm erf^{-1}}\left(\frac{1}{2}\right)$.\par
\begin{figure}
   %\centering
   \includegraphics[width=0.47\textwidth]{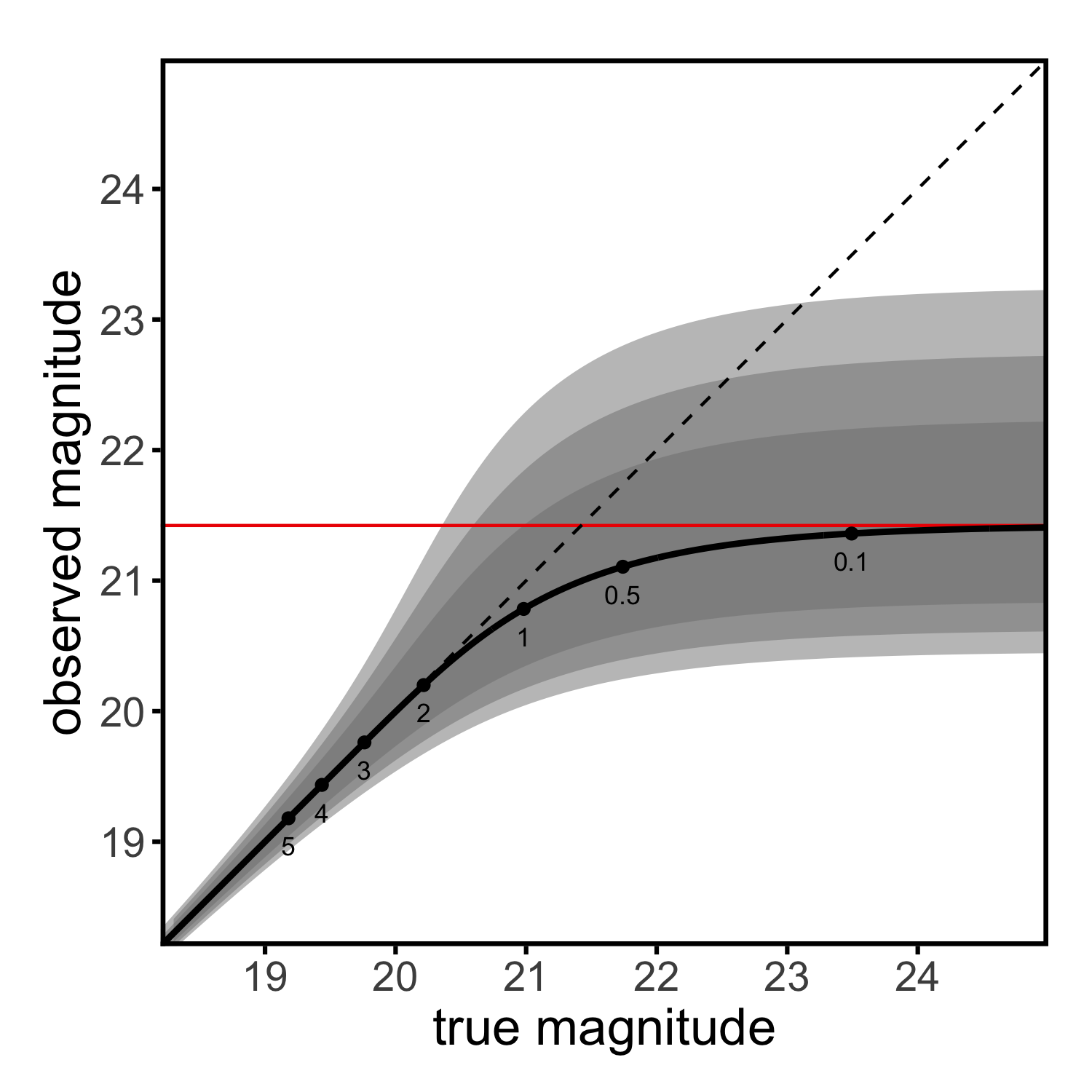}
   \caption{The relationship between the observed magnitude and the true magnitude for the example $\sigma_\phi = \sqrt{40^2 + \mu_\phi}$ and $zp = 25$. The solid black line shows the median of the magnitude distribution function (Eq. \ref{eq:newMag}), the grey shaded areas indicate the intervals between the 0.1 and 0.9, the $\Phi(-1)$ and $\Phi(1)$, and the 0.25 and the 0.75 quantiles of the distribution, respectively. The dots indicate signal to noise ratios between 5 and 0.1. The red line indicates the limiting magnitude according to Eq. (\ref{eq:limMag}).}
              \label{fig:dist1}
    \end{figure}

Measures of the uncertainty, such as confidence intervals between quantiles, are also linked to the pdf of the magnitude, and therefore they also converge towards limits as the magnitude distribution converges towards the limit distribution. Inevitably, also the uncertainties therefore are no longer realistic when the signal-to noise ratio approaches zero. Figure~\ref{fig:dist1} illustrates this effect for arbitrarily chosen values of $\sigma_\phi = \sqrt{40^2 + \mu_\phi}$, where the 40 represent a constant background and detector noise, and the dependency on the flux $\phi$ approximates the photon noise on the observation, and $zp = 25$ (the units are not of relevance for this numerical example, one might think for example of 40 photo-electrons per second and square metre, and a zero point of 25$^{\rm m}$). The solid black line shows the median of the distribution, i.e. the magnitude computed with Eq.~(\ref{eq:newMag}). The shaded regions indicate the intervals between some quantiles, as measurements for the uncertainty. The intermediate shaded region corresponds to the error interval as computed from Eqs.~(\ref{eq:newLow}) and (\ref{eq:newHigh}). Up to a signal-to-noise ratio of about 2, the observed magnitude agrees well with the true magnitude. For lower signal-to-noise ratios, the observed magnitude begin to become systematically too small, until all quantiles finally converge to the values for the limiting distribution, which is $m_{{\rm limit}}$ for the median.\par
This convergence behaviour is not a consequence of the use of magnitudes, but occurs in the flux domain as well. With decreasing signal-to-noise ratio, the distribution of \textit{observed} fluxes approaches a normal distribution, with zero mean and a standard deviation $\sigma$ which results from the sum of the background and detector noise. The distribution of \textit{true} fluxes however is different from this distribution. As negative true fluxes are impossible a-priori, the distribution of true fluxes approaches a half-normal distribution in the case of vanishing signal. The median of a half-normal distribution is $\sigma\, \sqrt{2}\, {\rm erf}^{-1}(1/2)$, and the limiting magnitude is the magnitude corresponding to this median, which is the magnitude for pure normally distributed random noise. Similarly, also the quantiles, and with it confidence regions, converge in the flux domain.\par
The convergence of the magnitudes to the limiting magnitude makes the latter an important quantity that characterises photometric observations. It might be useful to provide the limiting magnitude, as it provides some information at what magnitudes the estimated magnitudes become biased towards too small values (Often the bias is expressed with respect to the mean. But as we use the median as an estimator, here and in the following we only consider the bias of the median with respect to the true value).

   \begin{figure}
   %\centering
   \includegraphics[width=0.49\textwidth]{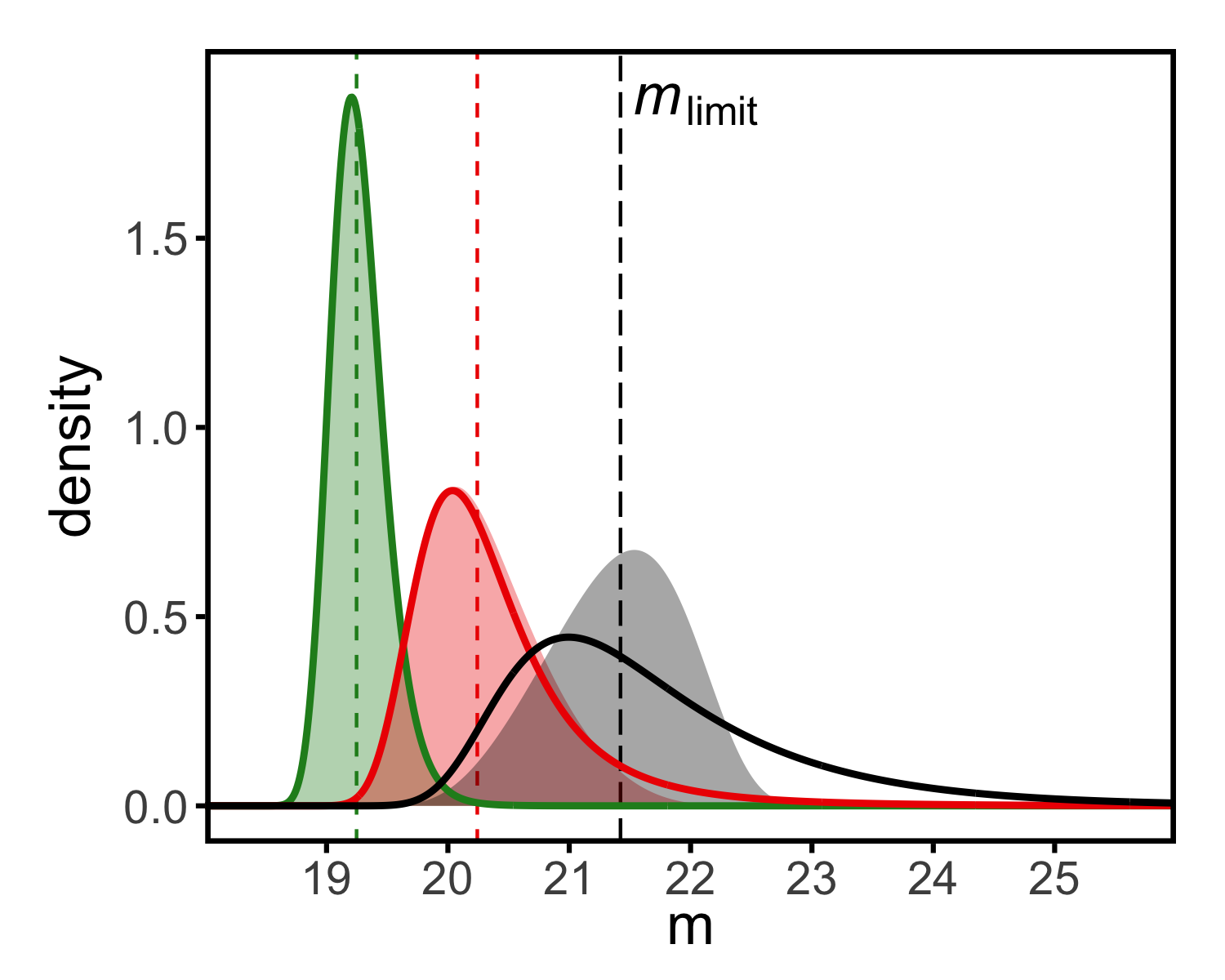}
   \caption{Solid lines show examples for pdfs of the true magnitude distributions, $f_{\boldsymbol{m}}(m)$, for different parameters. $\sigma_\phi = 40$ and $zp = 25$ in all cases, and the signal-to-noise ratios are 5 (green line), 2 (red), and 0 (black), respectively. The true magnitudes and the limiting magnitude are indicated by the dashed lines. The shaded regions show the pdfs of the median magnitude distributions for comparison.}
              \label{fig:magDistribution}
    \end{figure}

If the magnitudes are computed using Eq.~(\ref{eq:newMag}), which computes the most probable median true magnitude, the resulting magnitude distribution does not follow $f_{\boldsymbol{m}}(m)$, but becomes the median distribution function for the magnitude. As Eq.~(\ref{eq:newMag}) is transcendental in the observed flux, we cannot provide this distribution in a closed from. However, it is easily computed numerically from Eqs.~(\ref{eq:newMag}) and (\ref{eq:generalTransform}). Figure~\ref{fig:magDistribution} shows examples for the pdfs of the magnitude distribution and the median magnitude distribution for three different signal-to-noise levels. In all cases the median of the median magnitude distribution function coincides with the median of the magnitude distribution function, and also the median magnitude distribution becomes asymmetric. However, the median magnitude distribution function is far less tailed than the magnitude distribution function.

%===================================
\section{The colour distribution \label{sec:colours}}
%===================================

A frequent use of magnitudes is the computation of colours. It is therefore of interest to use the magnitude distribution from Eq.~(\ref{eq:magDistFunc}), and derive the colour distribution. The pdf of the sum of two independent random variables is given by the convolution of the pdfs of the two random variables \citep{Borovkov2013}. With Eq.~(\ref{eq:generalTransform}), the pdf of the the difference between two magnitudes $m_1 - m_2$, described by distribution functions with the parameters $\mu_1$, $\sigma_1$, $zp_1$, and $\mu_2$, $\sigma_2$, $zp_2$, respectively, and with the abbreviation $z \defeq m_1 - m_2$, becomes thus
\begin{equation}
f_{\boldsymbol{z}}(z) = f_{\boldsymbol{m_1}}(m_1) \ast f_{\boldsymbol{m_2}}(-m_2) \, .
\end{equation}
This pdf of the colour distribution can be expressed in a complicated but closed form. Using the abbreviations
\begin{eqnarray}
s_i & = & \frac{\mu_i}{\sigma_i}\;,\; i=1,2 \\
a & = & \left(\frac{10^{0.4\, zp_1}}{\sigma_1} \right)^2 + \left(\frac{10^{0.4\, (zp_2+z)}}{\sigma_2} \right)^2 \\
b & = & \frac{s_1}{\sigma_1} \, 10^{0.4\, zp_1} + \frac{s_2}{\sigma_2} \, 10^{0.4\, (zp_2 + z)}
\end{eqnarray}
one obtains
\begin{equation}
%\begin{multline}
\begin{split}
f_{\boldsymbol{z}}(z) = \frac{2}{\pi} \, \frac{{\rm ln}(10)}{2.5} \, \frac{1}{\sigma_1\, \sigma_2} \frac{1}{\left[1 + {\rm erf}\left(\frac{s_1}{\sqrt{2}}\right)\,\right]\, \left[1 + {\rm erf}\left(\frac{s_2}{\sqrt{2}}\right)\,\right]} \, \times \\
 10^{0.4\,(zp_1+zp_2+z)}\, \frac{1}{a}{\rm e}^{\frac{b^2}{2a}-\frac{s_1^2 + s_2^2}{2}}  \left({\rm e}^{-\frac{b^2}{2\, a}}  +  \sqrt{\frac{\pi}{2\,a}}\,b \left[ 1 + {\rm erf}\left(\frac{b}{\sqrt{2\, a}}\right)\, \right]\, \right)
 %\end{multline}
 \end{split}
\end{equation}
The function $f_{\boldsymbol{z}}(z)$ describes the statistical distribution of noise-affected colours at all signal-to-noise levels in both passbands involved. In limit of $s_1 = s_2 = 0$, i.e. in the limit of vanishing signal in both passbands, the pdf of this distribution simplifies to
\begin{equation}
\begin{split}
f_{\boldsymbol{z}}(z) = \frac{2}{\pi}\, \frac{{\rm ln}(10)}{2.5} \, &10^{0.4\, (zp_1+zp_2)} \, \times \\
 &\frac{10^{0.4\, z}}{\frac{1}{\rho}\left( 10^{0.4\, zp_1} \right)^2 + \rho\, \left( 10^{0.4\, zp_2} \right)^2 \, \left(10^{0.4\, z}\right)^2 } \label{eq:colLim}
\end{split}
\end{equation}
with $\rho = \sigma_1 / \sigma_2$ the ratio of the noise in the two passbands. This limit distribution has only one degree of freedom, its mean $\bar{z}$, which is identical to its median as the distribution is symmetric, and which is
\begin{equation}
\bar{z} = zp_1 - zp_2 - 2.5\cdot {\rm log_{10}}(\rho) \, ,
\end{equation}
and it has a standard deviation $\sigma$ of
\begin{equation}
\sigma = \frac{2.5}{{\rm ln}(10)} \, \frac{\pi}{2} \, .
\end{equation}
It describes the distribution of the colour of noise. While the median magnitude converges to the limiting magnitude given by Eq.~(\ref{eq:limMag}) for vanishing signal, the median colour converges to the limit colour $\bar{z}$ for vanishing signals in both passbands.\par
\begin{figure}
   %\centering
   \includegraphics[width=0.49\textwidth]{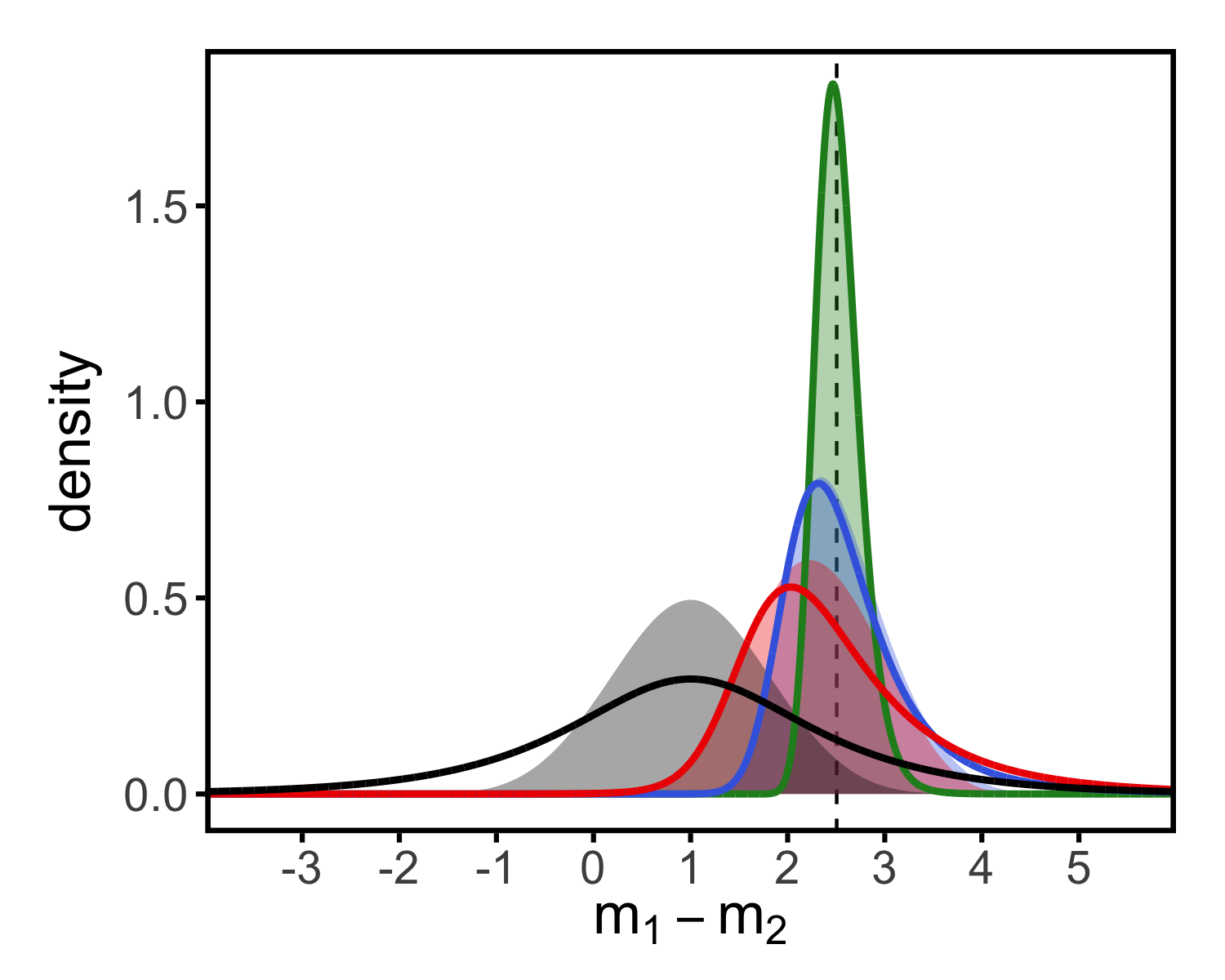}
   \caption{Solid lines show examples for pdfs of colour distributions, $f_{\boldsymbol{z}}(z \equiv m_1-m_2)$, for different parameters. $\rho = 1$ in all cases, and the signal-to-noise ratios in the two passbands are 5, 20 (green line), 2, 8 (blue), 1, 4 (red), and 0, 0 (black), respectively, and $zp_1 - zp_2 = 1$. The true colour is indicated by the dashed line. The shaded regions show the pdfs of the median colour distributions for comparison.}
              \label{fig:colourDistribution}
    \end{figure}

Figure \ref{fig:colourDistribution} shows four pdfs of colour distributions for the same true colour, but with different signal-to-noise levels in the two passbands. One can see how the colour distribution develops from a approximately symmetric distribution for high signal-to-noise values through asymmetric and biased distributions to the limit distribution of Eq.~(\ref{eq:colLim}), which again is symmetric and whose mean is solely depending on the difference in passband zero points and the logarithm of $\rho$.

If the magnitudes are computed using Eq.~(\ref{eq:newMag}), then the resulting colours are not distributed according to $f_{\boldsymbol{z}}(z)$, but according to the convolution of the pdfs of the mean magnitude distribution of one magnitude with the flipped one for the other passband. Again, as the median magnitude distribution can only be computed numerically, also the median colour distribution can only be computed numerically. The shaded regions in Fig.~\ref{fig:colourDistribution} show the pdfs of the median colour distributions in comparison to the ones of the colour distributions. Just as the median magnitude distributions, also the median colour distributions are far less tailed for low signal to noise ratios.

%===================================
\section{The distance distribution \label{sec:distances}}
%===================================

\subsection{Priors for the distance}

In the following we consider the problem of deriving the distance of an object from its parallax. Both quantities are, in the limit of small angles, related by Eq.~(\ref{eq:defDist}). We shall consider the possibility that the parallax $\varpi$ has a low signal-to-noise ratio. The error on the parallax we assume to be normally distributed, with a standard deviation $\sigma_\varpi$. In this case, we encounter a situation similar to the case of the flux, as also the parallax is a-priori a non-negative quantity.\par
The use of priors in the distance has been discussed extensively before, and priors on the distance are widely used. Following \cite{LutzKelker1973}, \cite{Bailer-Jones2015} introduced a prior that assumes a homogeneous volume distribution of stars, which results in a dependency of the prior on $D^2$. The motivation for this dependency of the distance prior on the distance is that within a sphere with a radius $R$, a constant volume density of stars, ${\rm d}N/{\rm d}V$, is proportional to the $1/V$, with $V$ the volume of the sphere. Using ${\rm d}V = {\rm d}r\, {\rm d}A$, with ${\rm d}A = r^2 \, {\rm sin}(\theta)\, {\rm d}\theta \, {\rm d}\varphi$ the surface element in spherical coordinates $(r,\, \theta,\, \varphi)$, one obtains for the radial density of sources
\begin{equation}
\frac{{\rm }d N}{{\rm d} r} \sim \frac{r^2}{R^3}\, {\rm sin}(\theta)\, {\rm d}\theta\, {\rm d}\varphi \quad . \label{eq:radialPrior}
\end{equation}
After integration over some solid angle, and normalisation, the probability distribution for the radial distribution of stars becomes
\begin{equation}
f_{\boldsymbol r}(r) = \frac{3}{R^3}\, r^2 \quad,
\end{equation}
which motivated the used of a dependency of the distance prior on the squared distance. Additional modifications have then been introduced to the prior to avoid a sudden cut at a fixed distance $R$, such as including an exponential decrease (resulting in the ``exponential decreasing space density'', EDSD, prior, \cite{Bailer-Jones2015}), or an exponential decrease with an additional modulation with a Gamma function \citep{Bailer-Jones2021}.\par
The assumption of a $D^2$-dependency as derived from Eq.~(\ref{eq:radialPrior}) is however not appropriate for the problem of finding a star's distance from a parallax. This dependency on distance solely holds if one considers a solid angle, which might include the entire celestial sphere or only the field of view covered by a single pixel during an observation. The transformation from parallax to distance however is not of such observational nature, but the purely geometrical problem of finding the position of a star along an imagined line running from the barycentre of the solar system through the star, given that one has measured the parallactic angle and its error. In this geometric problem, no solid angles are involved, and thus no dependency on the squared distance is involved. Introducing the prior that assumes an increasing probability with the squared distance therefore results in a strong overestimation of the distance. Problems with an overestimation of distances with priors that include a $D^2$-dependency have already been noted by \cite{Bailer-Jones2015}, who found that the mean or the median of the distance distribution resulting for the prior with a $D^2$ dependency are not suitable estimators, as they suggest much too large distances. This effect has been reduced by using the mode of the distribution, and in cases where the pdf of the distribution is bimodal, the mode at smaller distances. This mode is only weakly affected by the strong increase of the probability with increasing distance, and most of the effects of the prior is removed again.\par
Not introducing an inappropriate dependency on $D^2$ in the prior avoids such problems. In this work, we assume a prior that is zero for negative distances, and constant otherwise. This approach assigns to every point along the imagined line from the solar system barycentre through the star the same a-priori distance. With the same a-priori probability at any point in space, this approach corresponds to the assumption of a constant spatial density of stars, but leaves out the integration over some solid angle.\par
For a prior that is constant in distance, \cite{Bailer-Jones2015} reported the problem that the resulting pdf for the true distance cannot be normalised, and thus is actually not a pdf. This finding contributed to the introduction of a prior that truncates the probability distribution at large distances. The non-normalisable distribution however resulted from a computational inaccuracy. A constant prior results in the pdf for the parallax, given the distance, being equal to the pdf for the distance, given the parallax. Then, the pdf for the distance can be obtained from writing pdf for the parallax in terms of distance. In \cite{Bailer-Jones2015} this has been done by simply exchanging the variable $\varpi$ by $D$ in the normal distribution for the observed parallax. The true distance, which was fixed, became the variable, and the observed parallax becomes fixed. If however a change of variables takes place in a pdf, it as to be ensured that the probabilities resulting from the pdf are preserved, which in this case means the condition
\begin{equation}
\int\limits_{\varpi_1}^{\varpi_2}\, f_{\boldsymbol \varpi}(\varpi)\, {\rm d}\varpi = \int\limits_{D_2 = 1/\varpi_2}^{D_1 = 1/\varpi_1} f_{\boldsymbol D}(D) \, {\rm d}D
\end{equation}
needs to hold for all $\varpi_1, < \varpi_2$. This requirement is met when $f_{\boldsymbol D}(D)$ is derived from $f_{\boldsymbol \varpi}(\varpi)$ by changing the variable from $\varpi$ to $D$, and multiplying the result with $|{\rm d}\varpi / {\rm d}D| = D^{-2}$. This procedure corresponds to a substitution of a variable and is similar to the problem of the transition from a spectral energy distribution expressed in frequency to the the spectral energy distribution expressed in wavelength. Also in this case the two distributions are equal, but the transition from frequency $\nu$ to wavelength $\lambda$ needs to make sure that the integral over the energy distribution is preserved, requiring that the result obtained after replacing the $\nu$ by $c/\lambda$, with $c$ the speed of light, has to be multiplied with $|{\rm d}\nu / {\rm d}\lambda| = c/\lambda^2$ to obtain the correct version of the spectral energy distribution in wavelength.\par
We thus obtain the pdf for the true distance by changing from the variable $\varpi$ to $D$ in the normal distribution for the pdf of $\varpi$ and by multiplying the result with $D^{-2}$. We can now include the non-negativity of the distances by assuming a prior which is identical to zero for $D < 0$, and constant for $D\ge 0$. To ensure that this prior can be normalised, we can proceed in an analogous way as for the fluxes, we assume that the universe is finite, but much larger than all distances for which the likelihood for the distance is significantly different from zero. Multiplying with such a prior means one has to renormalise the pdf on the interval $[0,\infty]$, and we obtain the pdf for the true distance, $f_{\boldsymbol{D}}(D)$, as
\begin{equation}
f_{\boldsymbol{D}}(D) = \sqrt{\frac{2}{\pi}} \, \frac{1}{\sigma_\varpi} \, \frac{1}{1+{\rm erf}\left(\frac{\mu_\varpi}{\sqrt{2}\, \sigma_\varpi}\right)} \, \frac{1}{D^2} \, {\rm e}^{-\frac{\left( \frac{1}{D} -\mu_\varpi \right)^2}{2\, \sigma_\varpi^2}} \, . \label{eq:distDist}
\end{equation}
The same result for $f_{\boldsymbol D}(D)$ is obtained if a prior on the parallax which is zero for negative true parallaxes and constant else is used. In this case, we obtain, analogous to the case of fluxes, a truncated normal distribution for the pdf of the true parallax, which is then transformed to the pdf of the corresponding true distance by using Eq.~(\ref{eq:generalTransform}). In this case the prior is more elegantly normalisable, as the true parallax is, as an angle, defined over a finite interval. A constant prior on the parallax and on the distance, including only the condition of non-negativity, are thus equivalent. This may be understood as a constant prior intuitively means assuming nothing, and it does not matter on which quantity one explicitly assumes nothing, parallax or distance.\par
From Eq.~(\ref{eq:distDist}) the cumulative distribution function and the quantile function of the distance distribution follow as
\begin{equation}
F_{\boldsymbol{D}}(D) = \frac{1}{1+{\rm erf}\left( \frac{\mu_\varpi}{\sqrt{2}\, \sigma_\varpi} \right)} \, \left[ 1 - {\rm erf}\left( \frac{1}{\sqrt{2}\, \sigma_\varpi \, D} - \frac{\mu_\varpi}{\sqrt{2}\, \sigma_\varpi}  \right) \right]
\end{equation}
and
\begin{equation}
D(F_{\boldsymbol{D}}) = \frac{1}{\mu_\varpi + \sigma_\varpi \, C\left( \frac{\mu_\varpi}{\sigma_\varpi},F_{\boldsymbol{D}}\right)} \, . \label{eq:distQuant}
\end{equation}
This quantile function is in its structure analogous to the one we obtained for the magnitudes, and we may use as the estimator for the distance and its uncertainly, $D^{+\sigma^+}_{-\sigma^-}$, the simple expressions
\begin{eqnarray}
D&  = & \frac{1}{\varpi + \sigma_\varpi \, C\left(\frac{\varpi}{\sigma_\varpi},\, \frac{1}{2} \right)}\label{eq:newDist} \\ 
\sigma^-&  = & D - \frac{1}{\varpi + \sigma_\varpi \, C\left(\frac{\varpi}{\sigma_\varpi},\, \Phi(-1) \right)} \label{eq:newDLow}\\
\sigma^+ & =& \frac{1}{\varpi + \sigma_\varpi \, C\left(\frac{\varpi}{\sigma_\varpi},\, \Phi(1) \right)} - D \, , \label{eq:newDHigh}
\end{eqnarray}
with $\varpi$ and $\sigma_\varpi$ the observed parallax $\in \mathbb R$ and its uncertainty, respectively, and $C(x,p)$ from Eq.~(\ref{eq:correctionFunction}). For vanishing signal, the distance distribution converges to a limit distribution, which is obtained by setting $\mu_\varpi$ in Eq.~(\ref{eq:distDist}) to zero. The median of the limit distribution is
\begin{equation}
D_{\rm limit} = \frac{1}{\sigma_\varpi \, \sqrt{2}\, {\rm erf^{-1}}\left(\frac{1}{2}\right)}\, .
\end{equation}
As discussed already for the case of magnitudes in Sect.~\ref{sec:magnitudes}, also the uncertainties of the distance converge if the distance distribution converges to the limit distribution, and they thus become unrealistic as well.

\subsection{Comparison of distances for the different priors}

   \begin{figure}
   %\centering
   \includegraphics[width=0.49\textwidth]{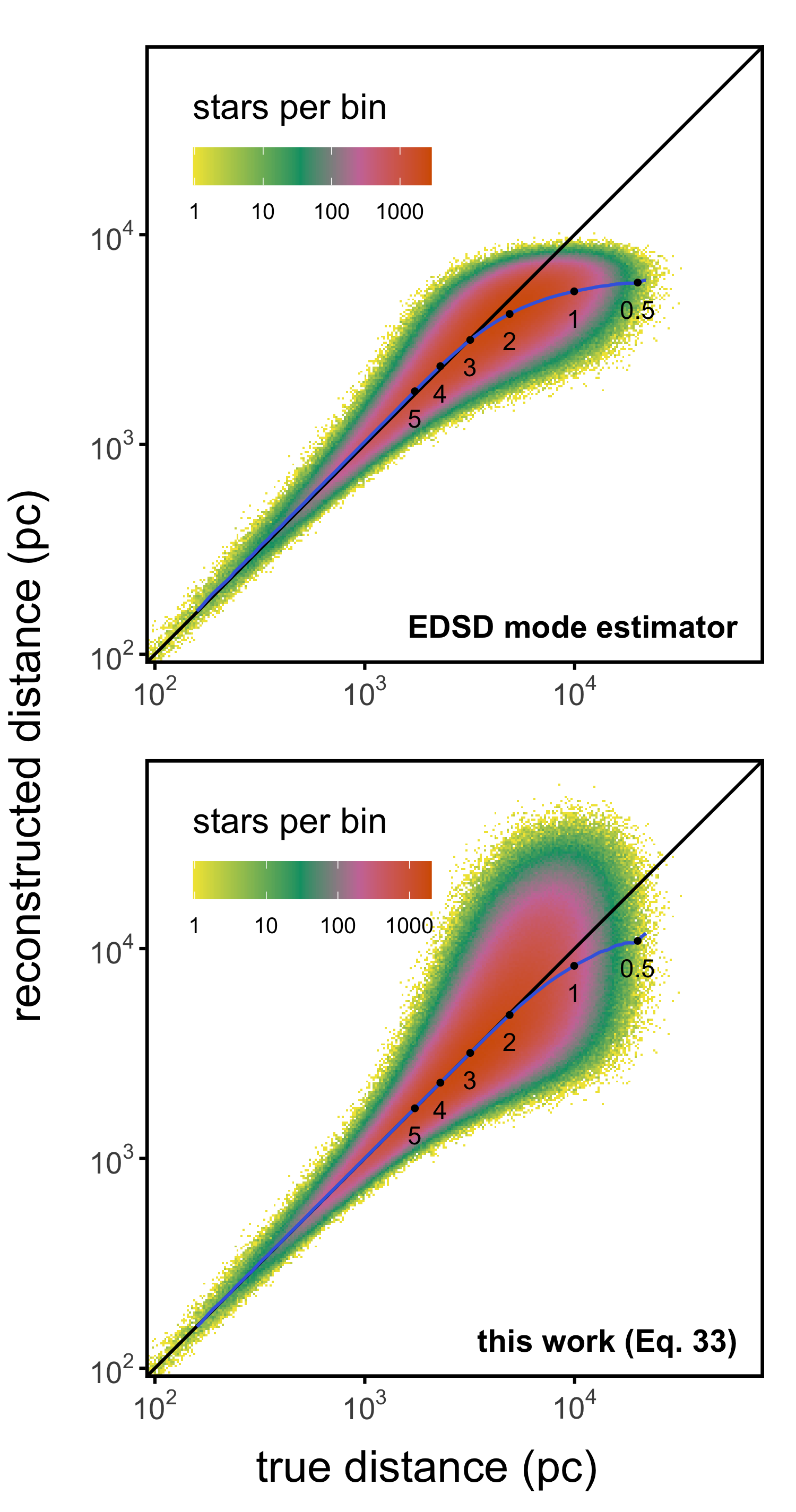}
   \caption{The relationship between the true distance and the reconstructed distance for five million simulated stars, using the EDSD mode estimator (top panel) and the estimator of this work (bottom panel). The blue lines show the running median of the reconstructed distance. Points indicate true signal-to-noise ratios between 5 and 0.5. For details see text.}
              \label{fig:priorComparison1}
    \end{figure}

   \begin{figure}
   %\centering
   \includegraphics[width=0.49\textwidth]{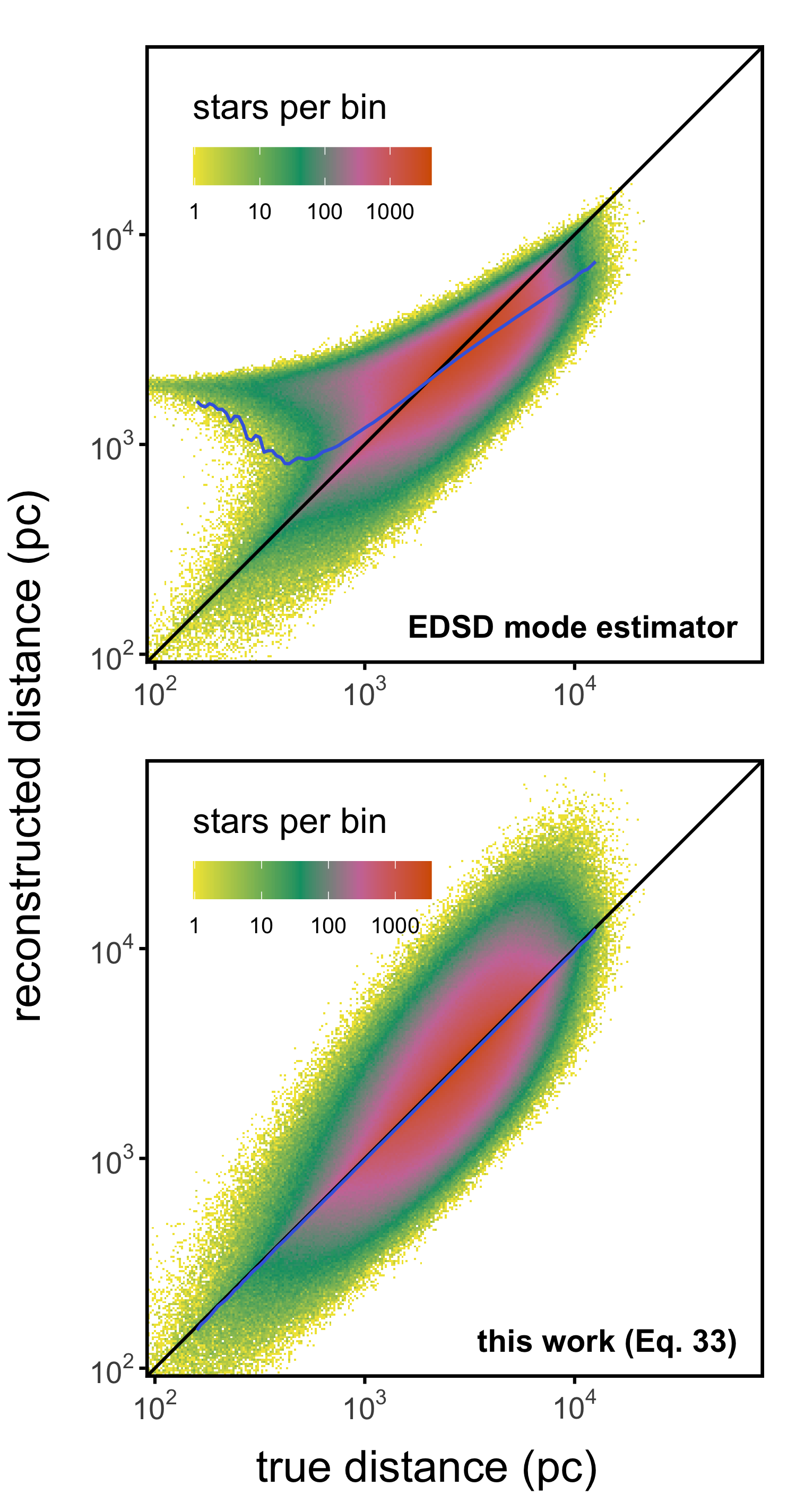}
   \caption{As figure~\ref{fig:priorComparison1}, but without error floor and for uniformly distributed fractional errors independent of distance.}
              \label{fig:priorComparison2}
    \end{figure}

   \begin{figure}
   %\centering
   \includegraphics[width=0.49\textwidth]{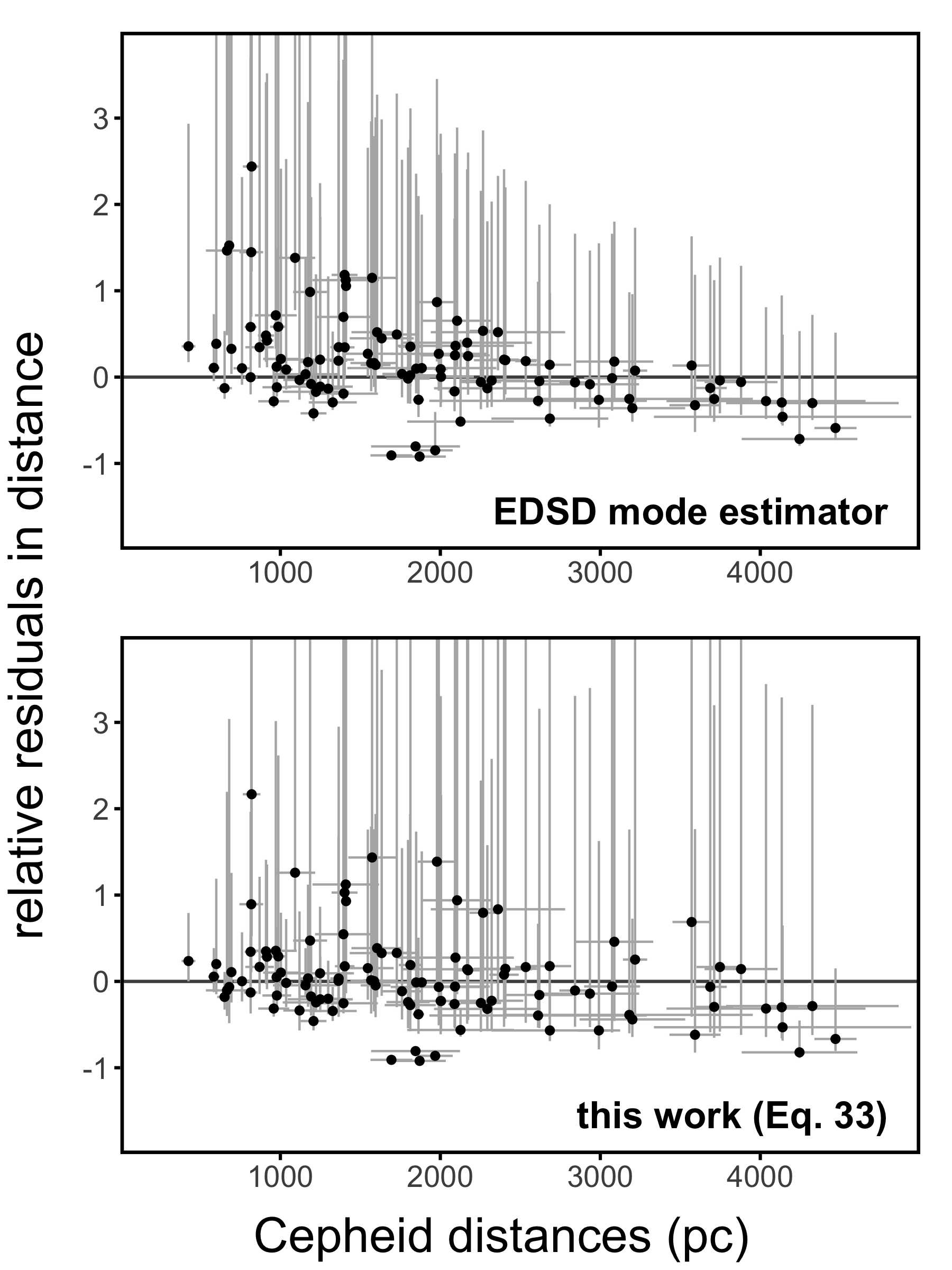}
   \caption{Comparison between the relative residuals in distance for Cepheids, using the EDSD mode estimator (top panel) and the estimator of this work (bottom panel). For details see text.}
              \label{fig:priorComparison3}
    \end{figure}

   \begin{figure}
   %\centering
   \includegraphics[width=0.49\textwidth]{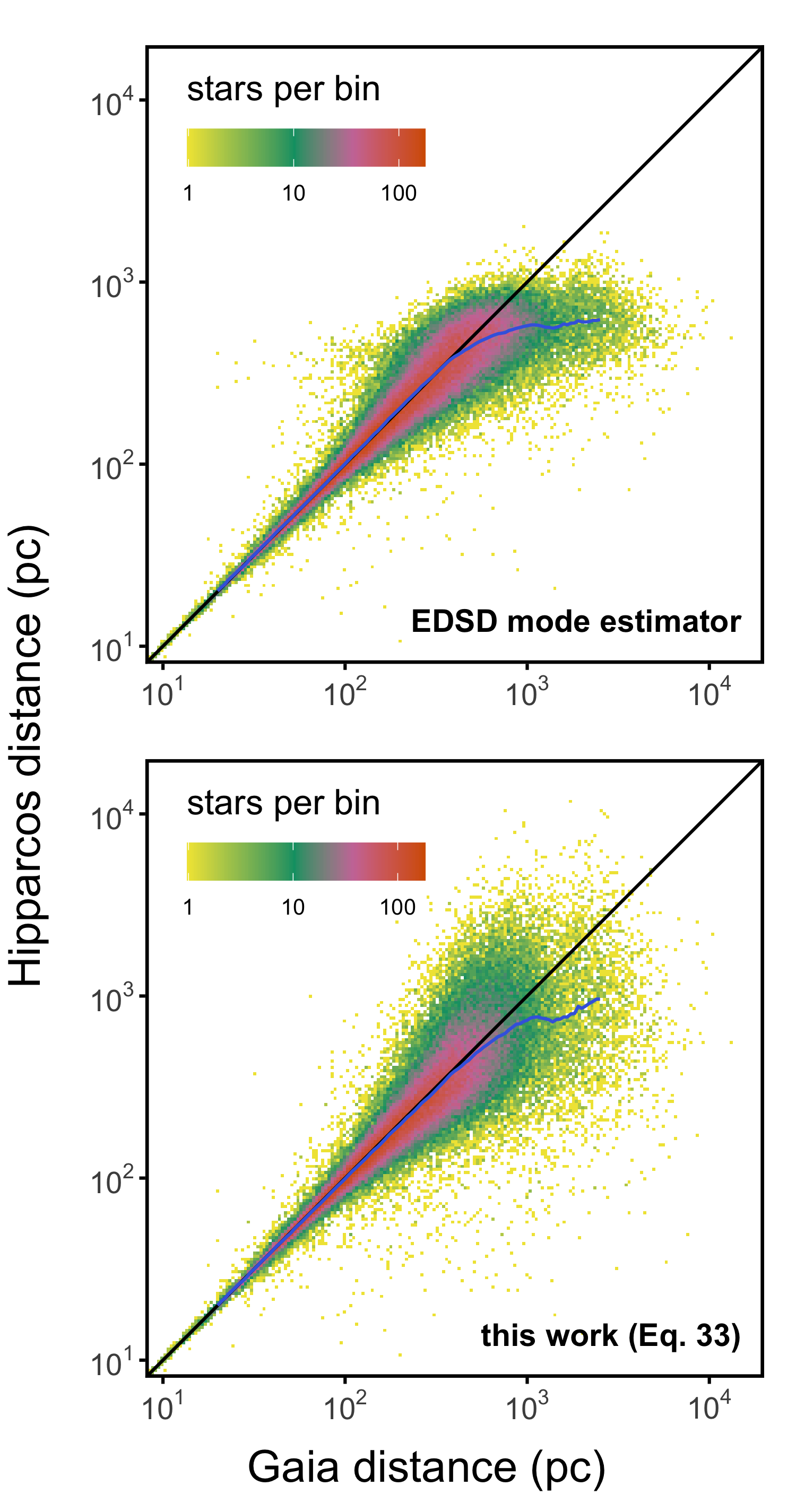}
   \caption{Comparison of the distances obtained from the Hipparcos parallaxes and from the Gaia parallaxes, for the EDSD mode estimator (upper panel) and the estimator of this work (lower panel). The blue lines show the running median. For details see text.}
              \label{fig:priorComparison4}
    \end{figure}

   \begin{figure*}
   \centering
   \includegraphics[width=0.85\textwidth]{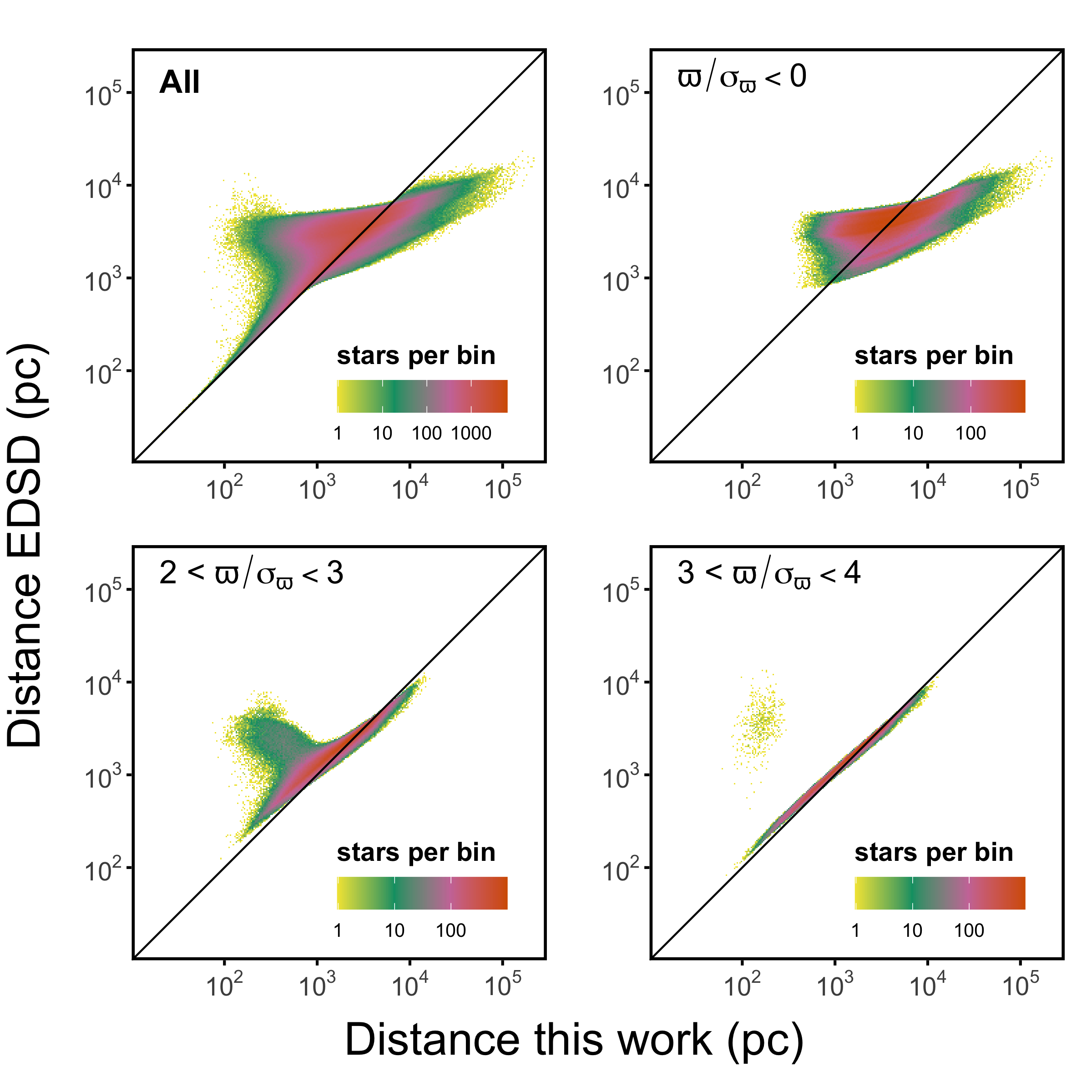}
   \caption{Comparison between the distance provided by \cite{Bailer-Jones2018}, using the EDSD mode estimator, and the distances obtained using Eq.~(\ref{eq:newDist}) of this work, for three million randomly selected sources. The upper left panel shows the distribution of all sources. The following panels show the distribution for those sources with an observational signal-to-noise of the parallax less than zero (upper right), for a signal-to-noise of the parallax between two and three (lower left), and between three and four (lower right).}
              \label{fig:priorComparison5}
    \end{figure*}

The pdf for the true distance derived in this work, assuming non-negativity and a constant space density of stars, is very different from previously suggested priors for this case. We therefore compare the distances obtained with the estimator from Eq.~(\ref{eq:newDist}), and from using a prior including a dependency on $D^2$ and a trunctation. For the comparison we use the least informative prior of its kind, the EDSD prior introduced by \cite{Bailer-Jones2015}. We use the mode estimator derived from this prior exactly as defined in this work. To compare the estimated distances with true distances for both priors, we do simulations. First, we draw true distances randomly, and to ensure that the scale length $L$ that occurs as an adjustable parameter in the EDSD prior is well chosen, we draw the true distances from the prior distribution. We then convert the true distances into true parallaxes. 
Then we compute observed parallaxes by adding normally distributed random noise to the true parallaxes. Finally, we compute reconstructed distances from the observed parallaxes, using mode estimator for the EDSD prior and the estimator of this work, given by Eq.~(\ref{eq:newDist}).\par
In a first simulation we use $L = 1.5 \times 10^3$ pc. For the standard deviation $\sigma_\varpi$ we use the quadratic sum of ten percent of the true parallax and a floor error of $10^{-4}$ arcseconds. With this choice for the observational error, the closest stars in our simulation have a signal-to-noise ratio of approximately ten, and the more distant simulated stars get, the smaller the signal-to-noise ratio of the parallax gets. Figure~\ref{fig:priorComparison1} shows the distribution of the reconstructed distances versus the true distances for five million simulated stars for both, using the EDSD mode estimator and the estimator of this work. The distribution of reconstructed distances using the EDSD estimator is narrower than the distribution using the estimator of this work. However, the reconstructed distances using the estimator of this work are less biased than for the EDSD estimator. Deviations between the median reconstructed distances and the true distances become visible at a true signal-to-noise ratio of approximately three for the EDSD estimator. For the estimator of this work, deviations become visible only from a true signal-to-noise ratio of two, and the deviations remain smaller for this estimator until the distant end of the distribution.\par
For a second test we use a configuration from \cite{Bailer-Jones2015}. We use a scale length $L$ of $10^3$~pc, set the floor error to zero, and use a fractional error, i.e. a noise-to-signal, uniformly random distributed in the interval $[0.01,1]$ independently of the true parallax. The result for five million simulated stars, for both estimators, are shown in Fig.~\ref{fig:priorComparison2}. Also in this case we observe that the distribution of the reconstructed distances using the EDSD estimator is narrower as compared to using the estimator of this work. And also in this case the median of the reconstructed distances is in better agreement with the true distances for the estimator of this work. The two branches in the distribution for the EDSD prior at small distances result from the discontinuity of the mode estimator for this prior. When the pdf of the distances goes from bimodal to unimodal, then, as discussed in \cite{Bailer-Jones2015}, the estimator jumps from the mode at smaller distances to the remaining mode, which is at much larger distances.\par
As seen in these two simulations, the detailed behaviour of the two estimators depend strongly on the assumptions made on the errors on the parallaxes, and both situations studied here might not be very realistic. In order to compare the estimators in realistic situations, we compare the two estimators also for two cases of real data. In the first case, we compare the results obtained for a number of Cepheid variables using the EDSD estimator as presented by \cite{Astra2016} with the results using the estimator of this work. The results are shown in Fig \ref{fig:priorComparison3}, showing the relative residuals in distance, $(D_{\rm estimated} - D_{\rm Cepheid})/D_{\rm Cepheid}$ against the Cepheid distance. The error bars indicate the range of relative residuals obtained by using the 0.05 and 0.95 quantiles, and the errors in the Cepheid distances have been scaled to the 90\% confidence interval. Also in this case, the distribution of the differences in distances resulting from the parallax and the Cepheid light curves is broader for the estimator of this work, as can be seen from the larger errors indicated. At the same time the residuals are better centred around zero. The unweighted mean and the standard deviation of the residuals are $0.16 \pm 0.54$ for the EDSD residuals, and $0.04 \pm 0.51$ for the residuals resulting from the estimator of this work.\par
In a second case we compare the distances computed from the parallaxes provided by the Hipparcos catalogue \citep{VanLeeuwen2007} with the distances obtained from the Gaia data release 3 catalogue \citep{Prusti2016, Valenari2023} for the same objects. Gaia parallaxes for the Hipparcos sources have a much higher signal-to-noise ratio, typically between 25 and 30 times higher, and the resulting distances are virtually independent from the prior used. As far as a comparison with Hipparcos is concerned, we may use the distances from Gaia parallaxes as estimates for the true distances. We thus use the best neighbour in the Gaia-Hipparcos crossmatch table, and we exclude the small fraction of sources with a signal-to-noise less than five in the Gaia DR3 catalogue. In order to ensure that we apply a good choice for the scale length $L$ for the Hipparcos catalogue, we keep $L$ a free parameter, and optimise it as to minimise the difference between the median of the reconstructed distances and the ``true'' distances, which is in this case the distances computed from the Gaia parallaxes. We did this optimisation in logarithmic space, and adopted a Hipparcos scale length of $L = 193$~pc. The result is shown in Fig.~\ref{fig:priorComparison4}. The distribution of Hipparcos distances is broader for the estimator of this work, as compared to using the EDSD mode estimator. The distribution obtained with the estimator of this work is however better centred on the true distances.\par
In all the comparisons we made, we always find that the distribution obtained with the estimator of this work being broader than for the EDSD estimator, but less biased.\par
Finally, we compare distances computed using the EDSD mode estimator and using Eq.~(\ref{eq:newDist}) of this work directly. For this purpose we select three million sources randomly from the table of distances provided by \cite{Bailer-Jones2018}, and compute the distances with the estimator of this work, using the corresponding parallaxes from the Gaia data release 2 catalogue \citep{Brown2018}, and applying the same parallax zero point correction as applied by \cite{Bailer-Jones2018}. The comparison is shown in Fig.~\ref{fig:priorComparison5}. The upper left panel of this figure shows the distribution for all the three million sources, and several structures are visible that depend on the observational signal-to-noise ratio. A spread of these structures is introduced by the use of different scale lengths in this data set, dependent on the galactic coordinates, and which vary in a complex way between about 335~pc and 2600~pc in this random data set, with a median of about 1473~pc. The upper right panel of Fig.~\ref{fig:priorComparison5} shows the distribution for sources with negative parallax only. The results for distances are different for this very low signal-to-noise regime, simply because the two priors have different limit distributions. For the EDSD mode estimator, the limiting distribution also changes with the scale length, and the distribution of scale lengths is producing the double feature seen in this panel. The bottom left panel compares distances for sources with a measured signal-to-noise ratio between two and three only, as this regime is the one where the largest systematic effects would be expected. For true signal-to-noise ratios between two and three, the estimator of this work is still virtually unbiased in simulations, while the EDSD mode estimator already shows some systematic deviation. And indeed the distances computed with the two estimators show a general systematic deviation, with larger distances for the EDSD estimator for nearer sources and smaller distances for more distant sources. The secondary branch of the distribution in this panel results from the discontinuity of the EDSD mode estimator, which results in a jump of the estimated distances towards too large values for some sources, and from those sources for which the median instead of the mode estimator is provided by \cite{Bailer-Jones2018}. Finally, the bottom right panel of Fig.~\ref{fig:priorComparison5} compares the distances for sources with a signal-to-noise ratio between three and four only. The overall deviation between the two estimators decreases with increasing signal-to-noise ratio. The isolated cluster of sources in this panel contains sources for which the median instead of the mode is provided as an estimator by \cite{Bailer-Jones2018}, resulting in too large distances. For further increasing signal-to-noise ratio, the systematic deviations reduce further. Four sources with a signal-to-noise ratio larger than five, the deviations reduce to about ten percent.

%===================================
\section{The absolute magnitude distribution \label{sec:absmagnitudes}}
%===================================

One use of distances is the combination with the magnitude, to obtain the absolute magnitude, $M$, which is given by
\begin{equation}
M = m - 5\cdot {\rm log_{10}}(D) + 5 \; ,
\end{equation}
where $D$ is the distance of the object measured in parsec. The pdf for $m$ we have already derived in Eq.~(\ref{eq:magDistFunc}), and the pdf for the distance distribution in Eq.~(\ref{eq:distDist}). In order to obtain the pdf for the absolute magnitude, we have first compute the pdf of the term $-5\cdot {\rm log_{10}}(1/\varpi)+5$ from the pdf for $\varpi$, using Eq.~(\ref{eq:generalTransform}), and then convolve the result with the pdf for the magnitude according to Eq.~(\ref{eq:magDistFunc}). We use the abbreviations
\begin{eqnarray}
s_\phi & = & \frac{\mu_\phi}{\sigma_\phi}\; , \; s_\varpi = \frac{\mu_\varpi}{\sigma_\varpi}\\
a ^\prime& = & \frac{\left(10^{0.4\, zp}\right)^2}{2\, \sigma_\phi^2}\\
b ^\prime & = & \frac{ \left( 10^{0.2\, M-1} \right)^2 }{2\, \sigma_\varpi^2} - \frac{ s_\phi }{ \sigma_\phi } \, 10^{0.4 \, zp}\\
c ^\prime & = & \frac{s_\varpi}{\sigma_\varpi} \, 10^{0.2\, M - 1} \,
\end{eqnarray}
with $\mu_\phi$ and $\sigma_\phi$ the mean and standard deviation of the flux, and $\mu_\varpi$ and $\sigma_\varpi$ the mean and standard deviation of the parallax. The distribution function for the absolute magnitude, for any level of signal-to-noise in both, flux and parallax, then becomes
\begin{equation}
\begin{split}
f_{\boldsymbol{M}}(M) = \sqrt{\frac{2}{\pi}}\, \frac{{\rm ln}(10)}{2.5} \frac{1}{\sigma_\phi\, \sigma_\varpi}\, \frac{1}{1+{\rm erf}\left(\frac{s_\phi}{\sqrt{2}}\right)}\, \frac{1}{1+{\rm erf}\left(\frac{s_\varpi}{\sqrt{2}}\right)}\, \times\\
{\rm e}^{-\frac{1}{2}\, \left(s_\phi^2 + s_\varpi^2\right)} \, 10^{0.4\, zp + 0.2\, M -1} \, \int\limits_0^\infty  x^2 \, {\rm e}^{-a ^\prime\, x^4 - b ^\prime\, x^2 + c ^\prime \, x}\, {\rm d}x \, . \label{eq:absMagDist}
\end{split}
\end{equation}
The function under the integral of this equation is zero for $x=0$, and it converges very quickly to zero for large arguments $x$. The resulting shape of this function is thus a rather narrow peak, which allows for an easy numerical evaluation of the integral.\par
\begin{figure}
   %\centering
   \includegraphics[width=0.49\textwidth]{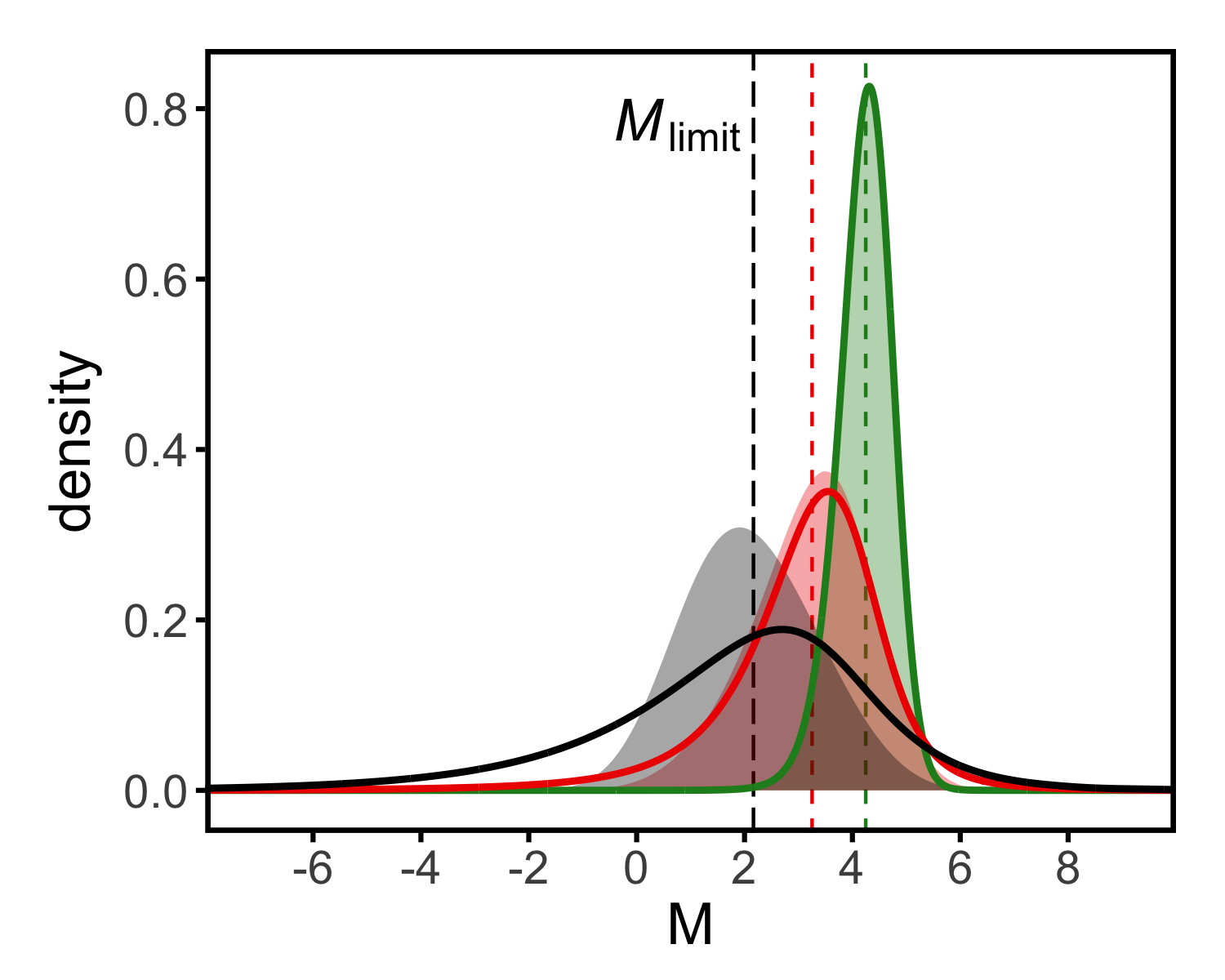}
   \caption{Solid lines show examples for pdfs of absolute magnitude distributions, $f_{\boldsymbol{M}}(M)$, for different parameters. The signal-to-noise ratios are 5 (green line), 2 (blue), and 0 (black), in both, flux and parallax. The true absolute magnitudes are indicated by the dashed lines. The black dashed line shows the limit absolute magnitude. The shaded regions show the pdfs of the median absolute magnitude distributions for comparison.}
              \label{fig:absMagDistribution}
    \end{figure}

Figure~\ref{fig:absMagDistribution} shows examples for the pdfs of absolute magnitude distributions for different parameters. As one would expect, for larger signal-to-noise ratios, the distribution function is centred on the true absolute magnitude, and almost symmetrical. For decreasing signal-to-noise ratios, the absolute magnitude distribution becomes increasingly asymmetric, and more and more tailed. In particular the tail towards smaller absolute magnitudes becomes very pronounced, making it increasingly likely to find a much smaller absolute magnitude than the true one. In the limit of vanishing signal-to-noise ratio, the distribution function converges to a very broad limiting distribution whose median, $M_{\rm limit}$, can be computed numerically starting from the pdf given by Eq.~(\ref{eq:absMagDist}).\par
We also compute the distribution function of the absolute magnitude when using Eq.~(\ref{eq:newMag}) for the computation of the magnitude, and Eq.~(\ref{eq:newDist}) for the computation of the distance. In this case, all steps, the computation of the pdfs of the magnitude and distance distribution functions, the conversion of the latter to the pdf for the expression $-5\cdot {\rm log_{10}}(1/\varpi)+5$, and the following convolution, have to be done numerically. The resulting pdfs of the median distribution function for the absolute magnitudes are also shown in Fig.~\ref{fig:absMagDistribution} for comparison. This distribution function also becomes asymmetric at low signal-to-noise ratios, but it is better centred on the true absolute magnitude, with much less pronounced tails. Eqs.~(\ref{eq:newMag}) and (\ref{eq:newDist}) are thus well suited also for the computation of the absolute magnitude.

%===================================
\section{Summary and practical considerations \label{sec:summary}}
%===================================

\subsection{Summary}
We have computed the pdf for the true magnitude given the observed flux, without any restrictions to the signal-to-noise ratio or non-negativity of the observed flux. This pdf becomes asymmetric at low signal-to-noise ratios, and it can be conveniently characterised by its quantiles. In particular the median provides a good measure for the magnitude at any signal-to-noise levels, and the $\Phi(-1)$ and $\Phi(1)$ quantiles for the uncertainly of the magnitude. Eqs. (\ref{eq:newMag}), (\ref{eq:newLow}), and (\ref{eq:newHigh}) can thus be used to compute the magnitude and its uncertainty for any observed flux and its error.\par
The corresponding median magnitude distribution, describing the statistical distribution of the estimated magnitudes, is far less tailed than the magnitude distribution, and it is thus avoiding extreme magnitude values. The median of the median magnitude distribution agrees well with the median of the magnitude distribution at all signal-to-noise levels. This median is in good agreement with the true magnitude down to signal-to-noise ratios as low as about two. For signal-to-noise levels converging towards zero, the magnitude distribution converges to a limit distribution. The median of this limit distribution, representing a limiting magnitude, and given by Eq.~(\ref{eq:limMag}), is a useful characteristic of the photometric observations. Observed magnitudes approaching the limiting magnitude will be biased towards too small values.\par
We derived the colour distribution for all signal-to-noise levels in either passband. This distribution is also strongly tailed, and it converges to a limit distribution whose only free parameter is its mean, depending on the the differences in the passband zero points and the logarithm of the noise ratio between the two passbands. If the magnitudes are computed with Eq.~(\ref{eq:newMag}), the resulting median colour distribution is also far less tailed, making the computation of magnitudes from Eq.~(\ref{eq:newMag}) suitable also for the computation of colours.\par
We applied the same formalism that we used for the computation of magnitudes to the problem of converting parallaxes into distances. We found that no $D^2$ dependency or truncation is required for the prior in distance, and we derived the pdf for the true distance, given the measured parallax and its uncertainty. This pdf includes only the non-negativity of the true parallax, and is constant for positive parallaxes. This approach is equivalent to using prior in on the true distances which is zero for negative distances, and constant otherwise. It is also consistent with assuming a constant spatial density of stars. Eqs. (\ref{eq:newDist}), (\ref{eq:newDLow}), and (\ref{eq:newDHigh}) provide an easy to use estimator for the true distance and its confidence interval at all signal-to-noise levels, including negative observed parallaxes. The closed form of the estimator allows for an easy and compact implementation, and it allows for a faster computation of distances from parallaxes than EDSD estimators, making it suitable for large data sets. It also does not involve any free parameters that need to be adjusted to particular data sets, but it can be readily applied to any parallax and its error. The resulting distribution for the estimated distance is broader than for previously suggested priors, but better centred on the true distance, and it provides reliable estimates down to a signal-to-noise ratio of about two. At lower signal-to-noise ratios, the estimated distances and their confidence intervals  become more and more biased, as the distance distribution is converging against the limit distribution for vanishing signal-to-noise.\par
Combining the computation of magnitudes using Eq.~(\ref{eq:newMag}) and the computation of the distance using Eq.~(\ref{eq:newDist}) in the computation of the absolute magnitude results in a distribution that is well centred on the true mean absolute magnitude, and which does not show strong tails. As the magnitudes, colours, and distances, also the absolute magnitudes converge to limiting distributions for vanishing signals. Their position can be characterised by the median of the limiting distributions.\par
While it is preferable to work with fluxes and parallaxes instead of magnitudes and distances wherever possible, the estimators presented in this work, derived from the least informative priors, provide a universally applicable solution when magnitudes or distances are desired, and when the construction of a more complex prior is not possible or desired.

\subsection{Practical considerations}

The prior on the flux used in this work introduces the non-negativity of the true flux, but makes no further assumptions. The resulting estimator for the magnitude can therefore be universally used, simply by replacing Eq.~(\ref{eq:def}) by Eq.~(\ref{eq:newMag}). Confidence intervals can be computed using the quantile function (Eq.~\ref{eq:quantiles}). Choosing the quantiles corresponding to the 1-$\sigma$ intervals of the standard normal distribution (Eqs. \ref{eq:newLow} and \ref{eq:newHigh}) results in a transition to the standard errors for high signal-to-noise ratios.\par
Similarly, the prior used in the computation of the distances only introduced the non-negativity of the true parallaxes. The constant value of the prior for parallaxes larger than zero is consistent with assuming a homogeneous density of stars in space. The resulting distance estimator can thus be universally used, by simply replacing Eq.~(\ref{eq:defDist}) with Eq.~(\ref{eq:newDist}). It might be possible to improve results in particular science cases by using a more adapted prior, or by combining parallaxes with additional information, as discussed by \cite{Luri2018}. Such adaptations depend however on the scientific case studied, while the distance estimator of this work can always be used and provides reliable results. As is the case for magnitudes, confidence intervals can be computed by using the quantile function (Eq. \ref{eq:distQuant}). The use of the quantiles corresponding to the 1-$\sigma$ intervals of the normal distribution (Eqs. \ref{eq:newDLow} and \ref{eq:newDHigh}) result in a confidence interval that transitions into the standard error of a normal distribution for high signal-to-noise ratios of the parallax.\par
The computation of the estimators and quantiles requires the evaluation of the correction function $C(x,p)$ (Eq.~\ref{eq:correctionFunction}). This function is defined for all $x \in \mathbb{R}$ and probabilities $p \in [0,1]$. The numerical evaluation might however become problematic if the argument $x$, which is the signal-to-noise ratio in the applications of this work, becomes too negative. In this case, the term ${\rm erf}(x/\sqrt{2})$ in the correction function might become numerically indistinguishable from $-1$, resulting in the evaluation of ${\rm erf}^{-1}(1)$, which is infinite, resulting in a numerical error. In double precision computations, this case occurs for values of $x$ below $-8$, i.e. if the measured value is more than eight standard deviations below zero. Since finding such a value with a normally distributed error and an a-priori positive true value is virtually impossible, it might be required to improve the input value or its estimated error should such an extreme case be encountered.

\begin{acknowledgements}
I would like to thank Josep Manel Carrasco, Claus Fabricius, Carme Jordi, and Xavier Luri for the helpful discussions.\\
This work was (partially) supported by the Spanish MICIN/AEI/10.13039/501100011033 and by ``ERDF A way of making Europe'' by the European Union through grants PID2021-122842OB-C21 and PID2021-125451NA-I00, the European Union ``Next Generation EU''/PRTR through grant CNS2022-135232, and the Institute of Cosmos Sciences University of Barcelona (ICCUB, Unidad de Excelencia `Mar{\'i}a de Maeztu') through grant CEX2019-000918-M.\\
This work has made use of data from the European Space Agency (ESA) mission
{\it Gaia} (\url{https://www.cosmos.esa.int/gaia}), processed by the {\it Gaia}
Data Processing and Analysis Consortium (DPAC,
\url{https://www.cosmos.esa.int/web/gaia/dpac/consortium}). Funding for the DPAC
has been provided by national institutions, in particular the institutions
participating in the {\it Gaia} Multilateral Agreement.
\end{acknowledgements}

\bibliographystyle{aa} % style aa.bst
\bibliography{Magnitudes} % your references Yourfile.bib

\begin{thebibliography}{15}
\expandafter\ifx\csname natexlab\endcsname\relax\def\natexlab#1{#1}\fi

\bibitem[{{Astraatmadja} \& {Bailer-Jones}(2016)}]{Astra2016}
{Astraatmadja}, T.~L. \& {Bailer-Jones}, C. A.~L. 2016, \apj, 833, 119

\bibitem[{Bailer-Jones(2015)}]{Bailer-Jones2015}
Bailer-Jones, C. A.~L. 2015, PASP, 127, 994

\bibitem[{{Bailer-Jones} {et~al.}(2021){Bailer-Jones}, {Rybizki}, {Fouesneau},
  {Demleitner}, \& {Andrae}}]{Bailer-Jones2021}
{Bailer-Jones}, C.~A.~L., {Rybizki}, J., {Fouesneau}, M., {Demleitner}, M., \&
  {Andrae}, R. 2021, \aj, 161, 147

\bibitem[{{Bailer-Jones} {et~al.}(2018){Bailer-Jones}, {Rybizki}, {Fouesneau},
  {Mantelet}, \& {Andrae}}]{Bailer-Jones2018}
{Bailer-Jones}, C.~A.~L., {Rybizki}, J., {Fouesneau}, M., {Mantelet}, G., \&
  {Andrae}, R. 2018, \aj, 156, 58

\bibitem[{{Borovkov}(2013)}]{Borovkov2013}
{Borovkov}, A.~A. 2013, {Probability Theory} (Springer)

\bibitem[{{Gaia Collaboration} {et~al.}(2018){Gaia Collaboration}, Brown,
  Vallenari, Prusti, De~Bruijne, Babusiaux, Bailer-Jones, Biermann, Evans,
  Eyer, Jansen, {et~al.}}]{Brown2018}
{Gaia Collaboration}, Brown, A.~G., Vallenari, A., {et~al.} 2018, \aap, 616, A1

\bibitem[{{Gaia Collaboration} {et~al.}(2016){Gaia Collaboration}, {Prusti},
  {de Bruijne}, {Brown}, {Vallenari}, {Babusiaux}, {Bailer-Jones}, {Bastian},
  {Biermann}, {Evans}, {Eyer}, {Jansen}, {Jordi}, {Klioner}, {Lammers},
  {Lindegren}, {Luri}, {Mignard}, {Milligan}, {Panem}, {Poinsignon},
  {Pourbaix}, {Randich}, {Sarri}, {Sartoretti}, {Siddiqui}, {Soubiran},
  {Valette}, {van Leeuwen}, {Walton}, {Aerts}, {Arenou}, {Cropper}, {Drimmel},
  {H{\o}g}, {Katz}, {Lattanzi}, {O'Mullane}, {Grebel}, {Holland}, {Huc},
  {Passot}, {Bramante}, {Cacciari}, {Casta{\~n}eda}, {Chaoul}, {Cheek}, {De
  Angeli}, {Fabricius}, {Guerra}, {Hern{\'a}ndez}, {Jean-Antoine-Piccolo},
  {Masana}, {Messineo}, {Mowlavi}, {Nienartowicz}, {Ord{\'o}{\~n}ez-Blanco},
  {Panuzzo}, {Portell}, {Richards}, {Riello}, {Seabroke}, {Tanga},
  {Th{\'e}venin}, {Torra}, {Els}, {Gracia-Abril}, {Comoretto},
  {Garcia-Reinaldos}, {Lock}, {Mercier}, {Altmann}, {Andrae}, {Astraatmadja},
  {Bellas-Velidis}, {Benson}, {Berthier}, {Blomme}, {Busso}, {Carry},
  {Cellino}, {Clementini}, {Cowell}, {Creevey}, {Cuypers}, {Davidson}, {De
  Ridder}, {de Torres}, {Delchambre}, {Dell'Oro}, {Ducourant}, {Fr{\'e}mat},
  {Garc{\'\i}a-Torres}, {Gosset}, {Halbwachs}, {Hambly}, {Harrison}, {Hauser},
  {Hestroffer}, {Hodgkin}, {Huckle}, {Hutton}, {Jasniewicz}, {Jordan},
  {Kontizas}, {Korn}, {Lanzafame}, {Manteiga}, {Moitinho}, {Muinonen},
  {Osinde}, {Pancino}, {Pauwels}, {Petit}, {Recio-Blanco}, {Robin}, {Sarro},
  {Siopis}, {Smith}, {Smith}, {Sozzetti}, {Thuillot}, {van Reeven}, {Viala},
  {Abbas}, {Abreu Aramburu}, {Accart}, {Aguado}, {Allan}, {Allasia},
  {Altavilla}, {{\'A}lvarez}, {Alves}, {Anderson}, {Andrei}, {Anglada Varela},
  {Antiche}, {Antoja}, {Ant{\'o}n}, {Arcay}, {Atzei}, {Ayache}, {Bach},
  {Baker}, {Balaguer-N{\'u}{\~n}ez}, {Barache}, {Barata}, {Barbier}, {Barblan},
  {Baroni}, {Barrado y Navascu{\'e}s}, {Barros}, {Barstow}, {Becciani},
  {Bellazzini}, {Bellei}, {Bello Garc{\'\i}a}, {Belokurov}, {Bendjoya},
  {Berihuete}, {Bianchi}, {Bienaym{\'e}}, {Billebaud}, {Blagorodnova},
  {Blanco-Cuaresma}, {Boch}, {Bombrun}, {Borrachero}, {Bouquillon}, {Bourda},
  {Bouy}, {Bragaglia}, {Breddels}, {Brouillet}, {Br{\"u}semeister},
  {Bucciarelli}, {Budnik}, {Burgess}, {Burgon}, {Burlacu}, {Busonero}, {Buzzi},
  {Caffau}, {Cambras}, {Campbell}, {Cancelliere}, {Cantat-Gaudin}, {Carlucci},
  {Carrasco}, {Castellani}, {Charlot}, {Charnas}, {Charvet}, {Chassat},
  {Chiavassa}, {Clotet}, {Cocozza}, {Collins}, {Collins}, \&
  {Costigan}}]{Prusti2016}
{Gaia Collaboration}, {Prusti}, T., {de Bruijne}, J.~H.~J., {et~al.} 2016,
  \aap, 595, A1

\bibitem[{{Gaia Collaboration} {et~al.}(2023){Gaia Collaboration}, {Vallenari},
  {Brown}, {Prusti}, {de Bruijne}, {Arenou}, {Babusiaux}, {Biermann},
  {Creevey}, {Ducourant}, {Evans}, {Eyer}, {Guerra}, {Hutton}, {Jordi},
  {Klioner}, {Lammers}, {Lindegren}, {Luri}, {Mignard}, {Panem}, {Pourbaix},
  {Randich}, {Sartoretti}, {Soubiran}, {Tanga}, {Walton}, {Bailer-Jones},
  {Bastian}, {Drimmel}, {Jansen}, {Katz}, {Lattanzi}, {van Leeuwen}, {Bakker},
  {Cacciari}, {Casta{\~n}eda}, {De Angeli}, {Fabricius}, {Fouesneau},
  {Fr{\'e}mat}, {Galluccio}, {Guerrier}, {Heiter}, {Masana}, {Messineo},
  {Mowlavi}, {Nicolas}, {Nienartowicz}, {Pailler}, {Panuzzo}, {Riclet}, {Roux},
  {Seabroke}, {Sordo}, {Th{\'e}venin}, {Gracia-Abril}, {Portell}, {Teyssier},
  {Altmann}, {Andrae}, {Audard}, {Bellas-Velidis}, {Benson}, {Berthier},
  {Blomme}, {Burgess}, {Busonero}, {Busso}, {C{\'a}novas}, {Carry}, {Cellino},
  {Cheek}, {Clementini}, {Damerdji}, {Davidson}, {de Teodoro}, {Nu{\~n}ez
  Campos}, {Delchambre}, {Dell'Oro}, {Esquej}, {Fern{\'a}ndez-Hern{\'a}ndez},
  {Fraile}, {Garabato}, {Garc{\'\i}a-Lario}, {Gosset}, {Haigron}, {Halbwachs},
  {Hambly}, {Harrison}, {Hern{\'a}ndez}, {Hestroffer}, {Hodgkin}, {Holl},
  {Jan{\ss}en}, {Jevardat de Fombelle}, {Jordan}, {Krone-Martins}, {Lanzafame},
  {L{\"o}ffler}, {Marchal}, {Marrese}, {Moitinho}, {Muinonen}, {Osborne},
  {Pancino}, {Pauwels}, {Recio-Blanco}, {Reyl{\'e}}, {Riello}, {Rimoldini},
  {Roegiers}, {Rybizki}, {Sarro}, {Siopis}, {Smith}, {Sozzetti}, {Utrilla},
  {van Leeuwen}, {Abbas}, {{\'A}brah{\'a}m}, {Abreu Aramburu}, {Aerts},
  {Aguado}, {Ajaj}, {Aldea-Montero}, {Altavilla}, {{\'A}lvarez}, {Alves},
  {Anders}, {Anderson}, {Anglada Varela}, {Antoja}, {Baines}, {Baker},
  {Balaguer-N{\'u}{\~n}ez}, {Balbinot}, {Balog}, {Barache}, {Barbato},
  {Barros}, {Barstow}, {Bartolom{\'e}}, {Bassilana}, {Bauchet}, {Becciani},
  {Bellazzini}, {Berihuete}, {Bernet}, {Bertone}, {Bianchi}, {Binnenfeld},
  {Blanco-Cuaresma}, {Blazere}, {Boch}, {Bombrun}, {Bossini}, {Bouquillon},
  {Bragaglia}, {Bramante}, {Breedt}, {Bressan}, {Brouillet}, {Brugaletta},
  {Bucciarelli}, {Burlacu}, {Butkevich}, {Buzzi}, {Caffau}, {Cancelliere},
  {Cantat-Gaudin}, {Carballo}, {Carlucci}, {Carnerero}, {Carrasco},
  {Casamiquela}, {Castellani}, {Castro-Ginard}, {Chaoul}, {Charlot}, {Chemin},
  {Chiaramida}, {Chiavassa}, {Chornay}, {Comoretto}, {Contursi}, {Cooper},
  {Cornez}, {Cowell}, {Crifo}, {Cropper}, {Crosta}, {Crowley}, {Dafonte},
  {Dapergolas}, {David}, {David}, {de Laverny}, {De Luise}, \& {De
  March}}]{Valenari2023}
{Gaia Collaboration}, {Vallenari}, A., {Brown}, A.~G.~A., {et~al.} 2023, \aap,
  674, A1

\bibitem[{Gradshteyn \& Ryzhik(2007)}]{Gradshteyn2007}
Gradshteyn, I.~S. \& Ryzhik, I.~M. 2007, Table of integrals, series, and
  products, seventh edn. (Elsevier/Academic Press, Amsterdam)

\bibitem[{{Hearnshaw}(1996)}]{Hearnshaw1996}
{Hearnshaw}, J.~B. 1996, {The Measurement of Starlight, Two Centuries of
  Astronomical Photometry} (Cambridge University Press)

\bibitem[{{Lupton} {et~al.}(1999){Lupton}, {Gunn}, \& {Szalay}}]{Lupton1999}
{Lupton}, R.~H., {Gunn}, J.~E., \& {Szalay}, A.~S. 1999, \aj, 118, 1406

\bibitem[{{Luri} {et~al.}(2018){Luri}, {Brown}, {Sarro}, {Arenou},
  {Bailer-Jones}, {Castro-Ginard}, {de Bruijne}, {Prusti}, {Babusiaux}, \&
  {Delgado}}]{Luri2018}
{Luri}, X., {Brown}, A.~G.~A., {Sarro}, L.~M., {et~al.} 2018, \aap, 616, A9

\bibitem[{{Lutz} \& {Kelker}(1973)}]{LutzKelker1973}
{Lutz}, T.~E. \& {Kelker}, D.~H. 1973, \pasp, 85, 573

\bibitem[{{Oke} \& {Gunn}(1983)}]{Oke1983}
{Oke}, J.~B. \& {Gunn}, J.~E. 1983, \apj, 266, 713

\bibitem[{{van Leeuwen}(2007)}]{VanLeeuwen2007}
{van Leeuwen}, F. 2007, \aap, 474, 653

\end{thebibliography}

\end{document}